\documentclass[12pt]{iopart}

\usepackage[utf8]{inputenc}
\usepackage{amssymb}
\expandafter\let\csname equation*\endcsname\relax
\expandafter\let\csname endequation*\endcsname\relax
\usepackage{latexsym,amsfonts,amsmath,amsthm,amsbsy,multirow,slashed,wasysym,textcomp,dsfont,comment,mathtools,mathrsfs}

\usepackage{tikzpagenodes}
\usepackage{adforn}
\usepackage{adforn}
\usepackage{tikz}
\usetikzlibrary{backgrounds}
\usetikzlibrary{patterns}
\usetikzlibrary{arrows.meta}
\usetikzlibrary{shapes}
\tikzstyle{bag} = [align=center]
\usetikzlibrary{decorations.pathmorphing}

\usepackage{graphicx}
\usepackage[font=small,labelfont=bf, width=0.8\linewidth]{caption}
\usepackage{cite}
\usepackage{hyperref}

\usepackage{etoolbox}

\makeatletter
\def\@mkboth#1#2{}
\newlength\appendixwidth
\preto\appendix{\addtocontents{toc}{\protect\patchl@section}}
\newcommand{\patchl@section}{%
  \settowidth{\appendixwidth}{\textbf{Appendix }}%
  \addtolength{\appendixwidth}{1.5em}%
  \patchcmd{\l@section}{1.5em}{\appendixwidth}{}{\ddt}%
}
\makeatother

\newcommand{\badat}{\begin{alignedat}}
\newcommand{\eadat}{\end{alignedat}}

\usepackage{xcolor}

\def\calA{\mathcal{A}}

\def\calN{\mathcal{N}}
\def\calO{\mathcal{O}}
\def\calP{\mathcal{P}}
\def\calS{\mathcal{S}}

\def\tcalA{\widetilde{\mathcal{A}}}

\def\tDelta{\widetilde \Delta}
\def\tJ{\widetilde J}

\def\bh{{\bar h}}
\def\bm{{\bar m}}
\def\bw{{\bar w}}
\def\bz{{\bar z}}

\numberwithin{equation}{section}
\begin{document}

\begin{flushright}
SAGEX-22-12\\
CPHT-RR016.032022\\
HU-EP-22/13\\
TCDMATH 22-02
\end{flushright}

\title[Soft Theorems and Celestial Amplitudes]{The SAGEX Review on Scattering Amplitudes \\ Chapter 11: Soft Theorems and Celestial Amplitudes}

\author{Tristan McLoughlin$^{1,4}$, Andrea Puhm$^2$ and Ana-Maria Raclariu$^3$}

\address{$^1$ School of Mathematics \& Hamilton Mathematics Institute,
		Trinity College Dublin, Ireland}
\address{$^2$ CPHT, CNRS, Ecole Polytechnique, IP Paris, F-91128 Palaiseau, France}
\address{$^3$ Perimeter Institute for Theoretical Physics,
31 Caroline Street North, Waterloo, Ontario, Canada N2L 2Y5}
\address{$^4$ Institut f\"ur Physik und IRIS Adlershof, Humboldt-Universit\"at zu Berlin, \\
  Zum Gro{\ss}en Windkanal 2, D-12489 Berlin, Germany}
\ead{$^1$tristan@maths.tcd.ie, $^2$andrea.puhm@polytechnique.edu, $^3$araclariu@perimeterinstitute.ca}
\vspace{10pt}

\begin{abstract}
The soft limits of scattering amplitudes have been extensively studied due to their essential role in the computation of physical observables in collider physics.
The universal factorisation that occurs in these kinematic limits has been shown to be related to conservation laws associated with asymptotic, or large, gauge symmetries. 
This connection has led to a deeper understanding of the symmetries of gauge and gravitational theories and to  a reformulation of scattering amplitudes in a basis of boost eigenstates which makes manifest the two-dimensional global conformal symmetry of the celestial sphere.
The recast, or \textit{celestial}, amplitudes possess many of the properties of conformal field theory  correlation functions which has suggested a path towards a holographic description of asymptotically flat spacetimes. In this review we consider these interconnected developments in our understanding of soft theorems, asymptotic symmetries and conformal field theory with a focus on the structure and symmetries of the celestial amplitudes and their holographic interpretation.  
\end{abstract}

%
%
%
%
%

 \pagebreak
 
\tableofcontents

\section{Introduction}

The surprising simplicity and hidden structure of scattering amplitudes motivates the search for new conceptual formulations, beyond standard quantum field theory, to make practical calculations more efficient and provide a gateway to new physics. The infrared (IR) behaviour of scattering amplitudes, which captures the long-distance dynamics, provides a rich source of such simplicity and structure. Understanding the origin of IR singularities, as due to the late- and early-time emission of massless particles, and their cancellation in physical observables has been of practical importance going back to the work of Bloch and Nordsieck \cite{PhysRev.52.54} and the KLN theorem \cite{Kinoshita:1962ur, Lee:1964is}. The universality of IR behaviour, in the sense that it does not depend on  details of the short distance physics in any given process, reflects its fundamental nature and, for the case of real radiation, is often referred to as a soft theorem. 

The observation that in gravity and gauge theories, soft theorems are Ward identities for asymptotic symmetries \cite{Strominger:2013lka,Strominger:2013jfa,He:2014laa,Kapec:2014opa,He:2014cra,He:2015zea,Kapec:2015ena,Campiglia:2015qka,Campiglia:2015kxa} has provided new impetus for the study of both. In particular, for asymptotically flat spacetimes, the residual diffeomorphisms which preserve the appropriate boundary conditions comprise an enhancement of the Poincar\'e group to include infinite classes of transformations called supertranslations and superrotations~\cite{Bondi:1962px,Sachs:1962wk, Sachs:1962zza,Barnich:2009se,Barnich:2010eb}. These additional symmetries have been ~\cite{Strominger:2014pwa,Pasterski:2015tva} related to gravitational memory effects
~\cite{zel1974radiation,christodoulou1991nonlinear} whose potential observability 
is under active investigation. Such connections between soft theorems, asymptotic symmetries and memory effects has led to important new insights into the physics of theories with massless particles. The connection between superrotations and soft theorems for gravitons led to the discovery of a Ward identity for a Virasoro symmetry acting on the null boundary of flat spacetime~\cite{Kapec:2014opa,Kapec:2016jld}, which provided evidence for the proposal that quantum gravity in asymptotically flat, four-dimensional spacetime is dual to a two-dimensional theory which is called ``celestial conformal field theory" (CCFT). 

The conformal properties of the theory become manifest upon transforming from the standard basis of plane waves to conformal primary wavefunctions~\cite{Pasterski:2016qvg,Pasterski:2017kqt}, which are boost eigenstates and on which the Lorentz group acts as the group of conformal transformations of the two-sphere at null infinity~\cite{deBoer:2003vf}. This reformulation of amplitudes in new variables, celestial amplitudes, provides an interpretation as CCFT correlation functions of operators labelled by their scaling dimensions and spin. This remarkable insight has led to the active investigation of the CCFT structure from both the bulk amplitude perspective and through applying known CFT methods. 

In this review we aim to provide an overview of developments in these areas and the connections between them. We start in section~\ref{sec:soft} with a review of soft theorems in gravity and gauge theories. We then briefly discuss asymptotic symmetries, their Ward identities and relation to soft theorems in section~\ref{sec:st-as}. This leads to the topic of celestial amplitudes in section~\ref{sec:celestial-amplitudes}, where we describe the basic holographic map and the properties of amplitudes in the conformal primary basis. In section~\ref{sec:Celestial-symmetries} we consider the symmetries of celestial amplitudes, the algebraic structure of these symmetries, and their implications. Finally we discuss some open questions and future directions in section~\ref{sec:open}.

\section{Soft Theorems}
\label{sec:soft}
 Perhaps the original example of universal IR behaviour in scattering amplitudes is the observation that the low energy limit of photon scattering by charged matter is given by the Thomson cross section which depends only on the charge and mass of the scatterer and not its spin or any details of its structure \cite{klein1929streuung}. Low \cite{Low:1954kd} and Gell-Mann and Goldberger \cite{Gell-Mann:1954wra} extended this result beyond the strict zero-frequency limit to the term linear in the photon frequency where dependence on the spin angular momentum appears. The explicit computations were done for spin-$\frac{1}{2}$ particles but the results, \cite{Low:1954kd} in particular, were based on gauge and Lorentz invariance and so could be argued to hold for any system with these symmetries \cite{pais1967low, bardakci1968low, Saito:1969lga}. 
\begin{figure}
	\centering
\begin{eqnarray}
\begin{array}{ccc}
\includegraphics[scale=1.25]{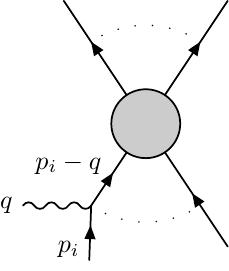}~~~& \includegraphics[scale=1.25]{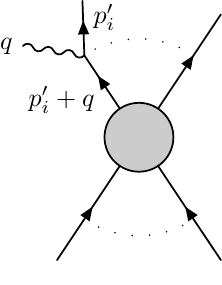}~~~~&~~~	\includegraphics[scale=1.25]{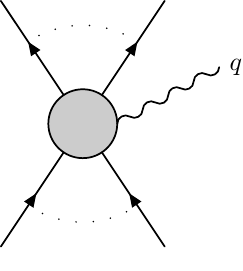}\\
(a) & (b) & (c)
\end{array}\nonumber
\end{eqnarray}
	\caption{Contributions to the photon emission amplitude.
	(a) and (b) give rise to soft poles while (c) contributes non-singular terms.}
	\label{fig:Soft}
\end{figure}

While such results pointed to the universal nature of soft limits, the constrained kinematics  obscured certain key features. Low's \cite{Low:1958sn} considerations of photon Bremsstrahlung radiation from spin-$0$ and spin-$\frac{1}{2}$ particles, generalised by Burnett and Kroll \cite{Burnett:1967km}, demonstrated the universal factorization of amplitudes with a soft-photon into soft factors and non-radiative lower point amplitudes. The Low-Burnett-Kroll (LBK) result depends on the fact that the amplitude with a photon contains two types of contributions: those for which the photon couples to an external line, which have a pole in the photon energy, and terms where the photon is coupled to an internal line, which do not have a singular contribution due to the off-shell nature of the virtual particles. The key roles played by Lorentz symmetry and on-shell gauge invariance motivated Weinberg \cite{Weinberg:1964ew, Weinberg:1965nx} and Gross and Jackiw \cite{Gross:1968in} to extend these results to graviton scattering.
 Jackiw \cite{Jackiw:1968zza} subsequently showed that the graviton soft theorems could also be derived from gauge invariance in a quite general fashion.

\subsection{Leading Soft Theorems}
As an example, let us consider a scattering process in QED with incoming electrons, with momenta $p_i$ and charges $Q_i$,  producing outgoing electrons with momenta $p'_i$ and charges $Q_i'$, interacting with an outgoing photon of momentum $q$ -- see Figure~(\ref{fig:Soft}). The contribution to the scattering amplitude $\mathcal{A}(p, p',q)$
from the photon coupling to an incoming leg, Figure~\ref{fig:Soft}.a, is given by\footnote{~Our conventions are those of \cite{Srednicki:2007qs} with metric signature $(-,+,+,+)$.
}
\begin{equation}
Q_i T(p_i-q) \frac{(-\slashed{p}_i+\slashed{q}+m){\slashed{\epsilon}}(q)}{(p_i-q)^2+m^2} u(p_i)
=-Q_i T(p_i-q)\Big[\frac{{\epsilon}(q)\cdot p_i+i {\epsilon}(q)_\mu q_\nu S^{\mu \nu} }{q\cdot p_i}\Big]u(p_i)
\end{equation}
where ${\epsilon}_\mu(q)$ is the outgoing photon polarisation vector and $S^{\mu \nu}=\frac{i}{4} [\gamma^\mu, \gamma^\nu]$ is the spin angular momentum generator acting on incoming electrons. $T(p_i-q)$ is the extrapolation, to the case with one leg shifted by $q$ to a non-physical momentum, of the elastic scattering amplitude for $n$ electrons, $\mathcal{A}(p, p')=T(p,p')u(p_i)$, with an incoming electron wavefunction, $u(p_i)$, stripped off.
 Other contributions come from vertices where the photon connects to outgoing electrons, Figure~(\ref{fig:Soft}.b), which have the same form as above but with the internal line having momentum $(p'_i+q)$ which results in an important sign, and contributions which are non-singular in the soft limit where the photon connects to an internal line, Figure~(\ref{fig:Soft}.c), which we denote $\epsilon_\mu(q)N^\mu(p,p',q)$. We can thus write the amplitude as 
\begin{eqnarray}
\label{eq:QED_amp}
\kern-60pt\mathcal{A}(p, p',q)&\kern-5pt=
\sum_{\rm{incoming}} -Q_i T(p_i-q)\frac{{\epsilon}(q)\cdot p_i+i \epsilon(q)_\mu q_\nu S^{\mu \nu} }{q\cdot p_i}u(p_i)
\\
&\kern-5pt+\sum_{\rm{outgoing}} Q_i'\bar{u}(p_i')\frac{{\epsilon}(q)\cdot p'_i+i{\epsilon}(q)_\mu q_\nu \bar{S}^{\mu \nu} }{q\cdot p_i'}\bar{T}(p'_i+q)
+{\epsilon}(q)_\mu N^\mu(p, p',q)~.\nonumber
\end{eqnarray}
In the soft limit, where $|q\cdot p_i|\leq |p_i \cdot p_j|$ for any external momenta $p_i$ and $p_j$, we have at leading order $T(p_i-q)u(p_i)\approx \mathcal{A}(p,p')$, and so dropping all non-singular terms, we find the leading soft theorem given as:
 \begin{eqnarray}
 \mathcal{A}(p, p',q)\approx \left[\sum_{\rm{outgoing}}Q'_i \frac{{\epsilon}(q)\cdot p'_i}{q\cdot p_i'}-\sum_{\rm{incoming}} Q_i\frac{{\epsilon}(q)\cdot p_i}{q\cdot p_i}\right]\mathcal{A}(p,p')~.
 \end{eqnarray}
 This result made use of the specifics of QED through the form of the vertex. However, Weinberg \cite{Weinberg:1964ew} showed that Lorentz invariance uniquely fixes the form of the soft photon coupling, for particles of charge $Q$ and arbitrary spin, to be proportional to the above result, i.e. $\propto Q \epsilon(q)\cdot p ~\delta_{\ell,\ell'} $, where $\ell$ and $\ell'$ are the initial and final helicities of the emitting particle. Weinberg also showed that  it is always possible to write the amplitude as a contraction of the polarisation vector and a Lorentz vector  $\mathcal{A}(p,p',q)={\epsilon}_\mu(q)A^\mu(p,p',q)$. The polarisation vector is not a true Lorentz vector and has a non-trivial little-group transformation, $\epsilon_\mu\to \epsilon_\mu+q_\mu$, also called an on-shell gauge transformation. Lorentz invariance of the scattering amplitude thus implies that it must vanish when ${\epsilon}_\mu$ is replaced by the momentum, $q_\mu A^\mu=0$. Consequently, by using the soft theorem, one sees that $\sum_{\rm{incoming}} Q_i=\sum_{\rm{outgoing}}Q'_i$ and so charge is conserved for arbitrary scattering processes in any Lorentz invariant theory of photons. 
 
 Weinberg generalised these considerations to massless particles of arbitrary integer spin $s$ where the amplitude can be written in terms of products of the spin-1 polarisation vectors $\epsilon_\mu(q)$ and $A^{\mu_1 \dots \mu_s}(q,p)$, a symmetric rank-$s$ Lorentz tensor,
 \begin{equation}
 \mathcal{A}(p,q)={\epsilon}_{\mu_1}(q)\dots{\epsilon}_{\mu_s}(q)A^{\mu_1 \dots \mu_s}(p,q)~,
 \end{equation}
 where for convenience we now denote all external legs, incoming and outgoing, by $p$.
Lorentz invariance implies that the amplitude must vanish when any of the polarisation vectors are replaced by the momentum $q^\mu$.
  For the spin-$2$  case of the graviton, the corresponding soft theorem is
 \begin{equation}
 \mathcal{A}(p,q)\approx \left[\sum_i \frac{\kappa_i}{2} \eta_i \frac{{\epsilon}_{\mu\nu}(q) p^\mu_ip_i^\nu}{q\cdot p_i}\right]\mathcal{A}(p)
 \end{equation}
 where ${\epsilon}_{\mu\nu}(q)=\epsilon_\mu(q){\epsilon}_\nu(q)$ is the polarisation tensor of the outgoing soft graviton written as a product of spin-one polarisation vectors, $\kappa_i$ is the ``gravitational charge" of the $i$-th particle and the sum goes over all external legs with $\eta_i=+1$ for outgoing particles and $-1$ for incoming. Repeating the argument from Lorentz invariance and using momentum conservation implies that all the $\kappa_i$ must be equal to $\kappa=\sqrt{32\pi G}$ and so, with the conventional definition of Newton's constant, that all particles have a gravitational mass equal to the inertial mass. Continuing to higher spin, $s>2$, the soft-limit is incompatible with Lorentz invariance which implies the impossibility of long-range higher-spin fields. 
 
 One important generalisation is to non-Abelian gauge fields where the soft limit of QCD amplitudes has a complicated and interesting structure which plays a key role in understanding the physics of high-energy colliders. However, at tree-level \cite{Berends:1987me, Berends:1988zn, Mangano:1987kp, Mangano:1990by}, the soft factorisation is very similar to that of QED but  we must replace the charge in the soft, also called the eikonal, factor by its matrix valued analogue $Q\frac{ {\epsilon} \cdot p}{q\cdot p}\to g T^a_R \frac{ {\epsilon} \cdot p}{q\cdot p}$ where $g$ is the gauge coupling and $T^a_R$ is the gauge group generator for the soft gluon with colour index $a$ acting on the external leg in representation $R$.

\subsection{Sub-leading Soft Theorems}
\label{sec:sub-leading-st}
 The gauge invariance arguments of \cite{Low:1958sn} and \cite{Burnett:1967km}, can be used to generalise the soft theorems to sub-leading orders.
 We consider a photon, now with polarisation vectors $\epsilon^\mu_\ell$ labelled by definite helicity $\ell=\pm$, and momentum $\delta q$, where $\delta\ll 1$ is an expansion parameter which defines the soft limit, in the scattering amplitude of $n$  particles with hard momenta $p_i$ and helicities $\ell_i$:
 \begin{equation}
 \label{eq:Sexp}
 \mathcal{A}_{\ell_1, \dots, \ell_n, \pm} ( p, \delta q)=\left[\sum_{a=0}^\infty \delta^{-1+a} S_\pm^{(a)}\right]\mathcal{A}_{\ell_1, \dots, \ell_n} ( p)~.
 \end{equation}
 Such an expansion is of course always possible, the power of the soft theorems lies in the fact that certain leading terms are universal, in that they are independent of the details of the amplitude. The soft operators $S_\pm^{(a)}$ are given as a sum of differential operators acting on each external leg and depending only on the charge, momentum and spin of that leg.\footnote{~One subtle point is the extrapolation of the amplitude on the right-hand side to non-physical momenta. The hard momenta satisfy $n+1$ particle conservation $\sum p_i=-q$ which is inappropriate for the $n$ particle amplitude. This is different than the prescription of \cite{Burnett:1967km} and thus leads to a different, but equivalent, form of the soft theorem.}  To continue with the  example of QED, using the condition $q_\mu A^\mu=0$ for the amplitude (\ref{eq:QED_amp}) implies that, to order $\mathcal{O}(\delta^0)$, we have 
 \begin{equation}
-q^\mu N_\mu(p,0)= \frac{1}{\delta} \sum_{i=1}^n  \eta_i Q_i\mathcal{A} ( p)+q^\mu \sum_{i=1}^n  Q_i \frac{\partial}{\partial p_i^\mu}~ \mathcal{A} ( p)~
 \end{equation}
 where we have suppressed the helicity indices on the amplitude and the derivative should be understood as not acting on wavefunction, $u(p_i)$ or $\bar{u}(p_i)$ as appropriate, in the amplitude. 
 Here we have used the assumption that $N^\mu(p,\delta q)$ has only analytical dependence on $q$, which is certainly valid at tree-level, so is independent of $q$ to this order. We can further use charge conservation to drop the leading $1/\delta$ contribution. This relation determines $N^\mu$ up to the addition of terms $v^\mu$ independently satisfying $q_\mu v^\mu=0$, however no such terms local in $q$ are possible. This expression for $N_\mu$ can be used to find that 
 \begin{equation}
 \kern-40pt A^\mu=\sum_{i=1}^n Q_i  \left[  \frac{\eta_ip^\mu_i}{\delta q\cdot p_i }+ \frac{q^\nu p^\mu_i }{q\cdot p_i}  \frac{\partial}{\partial p^\nu_i}-\frac{i q_\nu S_i^{\mu\nu}}{q\cdot p_i}-\frac{\partial}{\partial p_{i \mu}}\right]\mathcal{A}(p)+\mathcal{O}(\delta)~,
 \end{equation}
 where we can now view the derivatives as acting on the full amplitude but $S_i^{\mu\nu}$ should be then understood to be in the appropriate representation for the $i$-th external particle including the cancelling contribution from passing the spinor wavefunction through the action of the momentum derivatives. Defining the orbital 
 \begin{equation}
L_i^{\mu\nu}=i\left(p_i^\mu \frac{\partial}{\partial p_{i\nu}}-p_i^\nu\frac{\partial}{\partial p_{i\mu}}\right)
 \end{equation}
 and total angular momentum generators $J_i^{\mu \nu}=L_i^{\mu\nu}+S_i^{\mu\nu}$ for the $i$-th particle we see that the amplitude satisfies (\ref{eq:Sexp}) with the soft factors
 \begin{eqnarray}\label{eq:soft-photon-factors}
 S_\pm^{(0)}= \sum_{i=1}^n Q_i \eta_i \frac{{\epsilon}_{\pm\mu}(q) p_i^\mu}{q\cdot p_i}~, \quad {\rm and} \quad S_\pm^{(1)}=-i\sum_{i=1}^n Q_i  \frac{{\epsilon}_{\pm \mu}(q) q_\nu J_i^{\mu\nu}}{q\cdot p_i}~.
\end{eqnarray} 
This factorisation of the soft-limit through sub-leading order is often referred to as the LBK theorem. 
The same argument \cite{Jackiw:1968zza}, \cite{Bern:2014vva} (see also \cite{White:2014qia, Broedel:2014fsa}) can be used to derive the universal soft factors for gravitons. In fact, for the case of gravitons the gauge invariance is sufficient to constrain not only the sub-leading soft term but also the sub-sub-leading term and Einstein gravity amplitudes satisfy the expansion (\ref{eq:Sexp}) with
\begin{eqnarray}\label{eq:soft-graviton-factors}
& \kern-20pt S_\pm^{(0)}=  \frac{\kappa}{2}\sum_{i=1}^n \eta_i \frac{{\epsilon}_{\pm\mu \nu}(q) p^\mu_ip_i^\nu}{q\cdot p_i}~, \quad S_\pm^{(1)}=-i\frac{\kappa}{2} \sum_{i=1}^n \frac{{\epsilon}_{\pm\mu\nu}(q) p^\mu_i  q_\lambda J_i^{\nu\lambda}}{q\cdot p_i}~,\nonumber\\
& \quad \qquad   S_\pm^{(2)}=-\frac{\kappa}{4}\sum_{i=1}^n \eta_i \frac{{\epsilon}_{\pm\mu\nu}(q) q_\rho q_\sigma J_i^{\mu \rho}J_i^{\nu \sigma}}{q\cdot p_i}~,
\end{eqnarray}
where the graviton polarisation tensors can be written as ${\epsilon}_{\pm \mu\nu}(q)=\epsilon_{\pm\mu}(q){\epsilon}_{\pm \nu}(q)$. In these soft factors, the definition of the angular momentum operator depends on the spin of the particle but otherwise they are independent of the specifics of the hard particles. The sub-leading and sub-sub-leading graviton soft theorems involving $S^{(1)}$, $S^{(2)}$  were first derived, at tree-level and in four dimensions, in the seminal paper by Cachazo and Strominger \cite{Cachazo:2014fwa} by means of the BCFW recursion relations and using spinor-helicity formalism. The sub-leading soft behaviour had already been considered using next-to-eikonal methods~\cite{White:2011yy, Akhoury:2013yua} and, following \cite{Cachazo:2014fwa}, they were generalised to arbitrary dimensions in~\cite{Schwab:2014xua, AfkhamiJeddi:2014fia, Kalousios:2014uva, Zlotnikov:2014sva}, by use of the CHY formalism \cite{Cachazo:2013iea, Cachazo:2013hca} and to generic effective field theories with higher curvature interactions~\cite{Elvang:2016qvq, Laddha:2017ygw} and string theories~\cite{Bianchi:2014gla, Schwab:2014sla, DiVecchia:2015oba, Sen:2017xjn, Bianchi:2015lnw, Bianchi:2016tju, Higuchi:2018vyu}. Twistor approaches have also played an key role in the study of soft-limits, see for example~\cite{Geyer:2014lca, Adamo:2014yya, Adamo:2015fwa, Lipstein:2015rxa} and Chapter~6 \cite{Geyer:2022cey} of the SAGEX Review on Scattering Amplitudes \cite{Travaglini:2022uwo} for a discussion. 

In general, even at tree-level the sub-leading term in the photon soft theorem and the sub-sub-leading term for the graviton receive corrections which depend on the specifics of the theory through the structure of the three-point coupling of the soft particle to hard particles.  However, for the soft dilaton theorem, which is closely related to the soft graviton theorem in string theory, the universality extends to $\mathcal{O}(q)$ as the leading, sub-leading and sub-sub-leading \cite{Ademollo:1975pf, Shapiro:1975cz, DiVecchia:2015jaq, DiVecchia:2016amo} terms are the same in any string theory.  Continuing the expansion in $\delta$, for photons there is an infinite set of additional soft theorems, at least at tree-level, beyond sub-leading order, and for gravitons beyond sub-sub-leading order, but this requires projecting the amplitude onto a symmetric component to remove the undetermined contributions coming from $N^\mu$. For example, for photons this can be done using a symmetric tensor $\Omega_{\mu \nu_1 \dots \nu_m}$, such that $\Omega_{\mu \nu_1 \dots \nu_m}\partial_q^{\nu_1}\dots \partial_q^{\nu_m}A^{\mu}$ satisfies an additional soft theorem following from a Ward-Takahashi identity for large gauge transformations \cite{Hamada:2018vrw}, see also \cite{Li:2018gnc}.

 The sub-leading soft theorem for gluons has been discussed in the eikonal approach \cite{Laenen:2008gt, Laenen:2010uz} and can be directly found at tree-level by the replacement $Q\to g T^a_R$ in the photon soft factors. It was also derived by using the BCFW method in \cite{Casali:2014xpa} where it was given in terms of colour ordered amplitudes, $\mathcal{A}^{\rm c.o.}$, and using the spinor-helicity formalism which is very convenient for massless scattering.  The soft limit of an $(n+1)-$point colour ordered amplitude with a soft gluon of momentum $\delta q$ and helicity $\ell$ can be written as  
\begin{equation}
\mathcal{A}^{\rm c.o.}_{\ell_1,\dots, \ell_n,\ell}(p_1,\dots,p_{n}, \delta q)=\left[\frac{1}{\delta} S_\ell^{(0)}+S_\ell^{(1)}\right]\mathcal{A}^{\rm c.o.}_{\ell_1,\dots, \ell_n}(p_1,\dots, p_n)+\mathcal{O}(\delta)~,
\end{equation}
with the soft factors now being a difference of terms acting on pairs of legs adjacent to the soft gluon which are
\footnote{The spinor brackets are defined in terms of the decomposition of on-shell momenta, $p_{a\dot a}=-\lambda_a\bar{\lambda}_{\dot a}$, and are given by $\langle ij\rangle= \epsilon^{\dot a \dot b}\bar{\lambda}_{i\dot b}\bar{\lambda}_{j\dot a}$, $[ij]=\epsilon^{a b}\lambda_{i a}\lambda_{j b}$~.}
\begin{equation}
S_+^{(0)}=\frac{\langle \mu 1\rangle}{\langle q 1 \rangle \langle q \mu\rangle} -\frac{\langle \mu n\rangle}{\langle q n \rangle \langle q \mu \rangle }=\frac{\langle 1 n\rangle}{\langle 1 q \rangle \langle q n\rangle}~,~~~
S_+^{(1)}=\frac{1}{\langle n q\rangle} {\lambda}_q^{a} \frac{\partial}{\partial {\lambda}_n^{ a}} - \frac{1}{\langle 1 q\rangle} {\lambda}_q^{ a} \frac{\partial}{\partial {\lambda}_1^{a}}~,
\end{equation}
where we have restricted to the case of a positive helicity soft gluon and $|\mu\rangle$ is an arbitrary reference spinor. Open string amplitudes describing gluons and massive external states were analyzed in \cite{Schwab:2014fia, Bianchi:2015yta} which found the leading and sub-leading terms were the same as in Yang-Mills theories. 

The spinor-helicity formalism is also useful for expressing the universal factorisation properties of gluon amplitudes in collinear limits \cite{Berends:1987me, Mangano:1990by, Bern:1994zx}, see also the SAGEX review Chapter~1 \cite{Brandhuber:2022qbk}.  At tree-level, when two adjacent outgoing external legs with helicities $\ell_i$ and $\ell_{j}$ become collinear, so that $p_i\approx \alpha  p_P$, $p_{j}\approx (1-\alpha)p_P$, the colour ordered amplitude can be factorised as
\begin{equation}
\mathcal{A}^{\rm c.o.}_{\dots,\ell_i,\ell_j,\dots}(\dots, p_{i}, p_{j},\dots)\approx \sum_{\ell=\pm 1} \text{Split}_{\ell_i,\ell_j}^{\ell}(p_i, p_j)\mathcal{A}^{\rm c.o.}_{\dots,\ell,\dots}(\dots, p_{P}, \dots)~,
\end{equation}
where the sum is over the helicity, $\ell$, of the intermediate particle of momentum $p_P$ and the universal splitting amplitudes are 
\begin{eqnarray}
\label{eq:split}
& \kern-0pt\text{Split}^{+1}_{-1,-1}(p_i,p_j)=0~,~~\text{Split}^{+1}_{+1,+1}(p_i, p_j)=\frac{1}{\sqrt{\alpha(1-\alpha)}\langle{ij}\rangle}~,\\
&\kern-20pt\text{Split}^{-1}_{+1,-1}(p_i, p_j)=\frac{(1-\alpha)^2}{\sqrt{\alpha(1-\alpha)}\langle{ij}\rangle}~,~~~\text{Split}^{+1}_{+1,-1}(p_i, p_j)=\frac{-\alpha^2}{\sqrt{\alpha(1-\alpha)}[ij]}~.\nonumber
\end{eqnarray}
These gluon splitting amplitudes can be used, via the KLT relations \cite{Kawai:1985xq}, to find the universal tree-level behaviour of graviton amplitudes in the collinear limit \cite{Bern:1998sc, Bern:1998sv}. For two collinear gravitons with momenta $p_i$ and $p_j$, parameterised as in the gluon case, and with helicities $\ell_i+\bar{\ell}_i$, $\ell_j+\bar{\ell}_j$, the splitting amplitude into a graviton of helicity $\ell+\bar{\ell}$ is given by
\begin{eqnarray}
& \kern-0pt\text{Split}^{\ell+\bar{\ell}}_{\ell_i+\bar{\ell}_i,\ell_j+\bar{\ell}_j}(p_i,p_j)=-\frac{\kappa}{2}\langle ij \rangle[ji] \text{Split}^{\ell}_{\ell_i,\ell_j}(p_i,p_j)\text{Split}^{\bar{\ell}}_{\bar{\ell}_j,\bar{\ell}_i}(p_j,p_i)~.
\end{eqnarray}
The collinear limit is non-singular for gravitons, for example 
\begin{equation}
\text{Split}^{+2}_{+2,+2}(p_i,p_j)=-\frac{\kappa}{2\alpha(1-\alpha)}\frac{[ij]}{\langle ij\rangle}~,
\end{equation}
 but the universal terms can be identified by the non-vanishing phase they acquire as $p_i$ and $p_j$ are rotated about $p_P$. Alternatively, in $(+,+,-,-)$ signature the spinor brackets $[ij]$ and $\langle ij\rangle$ are independent and, for example, the universal term with $\text{Split}^{+2}_{+2,+2}$ can be seen to be leading in a  $\langle ij \rangle \to 0$ expansion with $[ij]$ held fixed.

\subsection{Loop Corrections}
The leading soft theorem for photons is exact in QED and has no loop corrections, however in general the same is not true for the sub-leading terms. As noted by Del Duca \cite{DelDuca:1990gz}, the LBK theorem only holds when the energy of the soft photon satisfies $\delta q^0\ll m^2/E$, where $m$ is the lightest mass of the external hard particles and $E$ is the energy scale characterising the scattering process, e.g. the center of mass energy. This is due to non-local terms, for example  poles like $\frac{1}{\delta p \cdot q}$, in  $N_\mu$, which cannot be expanded in small $\delta$. Such contributions can arise from regions of loop integration where the loop momentum~$l$ becomes collinear with an external leg with mass $m$. In this case the loop momentum has small virtuality $l^2 =\mathcal{O}(m^2)$ but has large individual components so that $l\cdot \delta q=\mathcal{O}(E \delta q^0)$. As a result, propagators in loop diagrams of the type
\begin{eqnarray}
\vcenter{\hbox{\includegraphics[scale=.75]{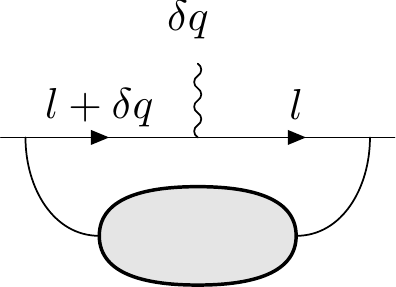}}}\Rightarrow \frac{1}{ (l+\delta q)^2+m^2}\approx \frac{1}{(l^2+m^2)+2\delta  l\cdot q}
\end{eqnarray}
can only be expanded when $\delta q^0\ll m^2/E$ and consequently for theories with massless particles, or in the high-energy limit with $E\to \infty$, the LBK theorem will fail. Loop corrections to soft-limits in non-Abelian gauge theory have been studied in detail \cite{Bern:1995ix, Bern:1998sc, Bern:1999ry, Kosower:1999rx,Catani:2000pi, Kosower:2003cz}, but remains an active topic with the full colour two-loop soft factor only computed relatively recently \cite{Duhr:2013msa, Li:2013lsa, Dixon:2019lnw}\footnote{See also the SAGEX review Chapter 12 \cite{White:2022wbr} for a discussion of loop corrections to sub-leading soft behaviour in gauge theory and further references to the literature including approaches within the context of Soft-Collinear Effective Theory (SCET) \cite{Bauer:2000ew, Bauer:2000yr, Bauer:2001ct, Bauer:2001yt} which provides a systematic approach to soft theorems in gauge theory, e.g. \cite{Larkoski:2014bxa}, and gravity \cite{Beneke:2012xa, Okui:2017all, Beneke:2021aip}.}.

Loop corrections to graviton soft theorems were discussed in \cite{Bern:2014oka, He:2014bga} based on the extensive earlier work in non-Abelian gauge theory. Theories with massless particles have IR singularities in the S-matrix which are crucial in understanding the loop level soft behaviour. For example, the one-loop $n$-graviton amplitude computed in dimensional regularisation, with IR regulator ${\varepsilon}=(4-D)/2$ and mass scale $\mu$, has a singular part proportional to the tree-level amplitude \cite{Dunbar:1995ed, Naculich:2011ry, Akhoury:2011kq}, \cite{Bern:2014oka} 
\begin{equation}
 \mathcal{A}^{1-{\rm loop}}=-\frac{\kappa^2}{32\pi^2 {\varepsilon} }\sum_{i<j}^n \eta_i \eta_j p_i\cdot p_j \log\left(-\frac{2 p_i\cdot p_j}{\mu^2}\right)\mathcal{A}^{\rm tree}+\mathcal{O}({\varepsilon}^0)~,
\end{equation}
where the sum is over all pairs of external momenta. If we were to take a naive soft-limit with an uncorrected soft-factor there would be a mismatch between the singular terms on the left and right. As a result the soft factor must receive loop corrections with singular terms 
\begin{equation}
\left. \mathcal{A}^{1-{\rm loop}}(p,\delta q)
 \right|_{\rm{ sing}}
 =\sum_{a} \delta^{a-1} \left.\right[  S^{(a)} \mathcal{A}^{1-\rm{loop}}(p)+ S^{(a), 1-\rm{loop}} \mathcal{A}(p)\left] \right|_{\rm{ sing}}~,
\end{equation}
and, potentially, additional finite terms. The singular corrections were computed in \cite{Bern:2014oka,Bern:2014vva}, see also \cite{Broedel:2014bza}, where it was shown that the leading soft-graviton factor is uncorrected at any loop order, while the sub-leading factor receives only one-loop corrections and sub-sub-leading terms are corrected at one and two-loops in perturbative gravity.

An alternative approach was followed in \cite{Cachazo:2014dia} where the soft-limit is taken before the expansion in ${\varepsilon}$. Unlike the standard approach, it is not known how to make this prescription consistent with the physical requirement that divergences cancel in observables such as cross-sections, but it does lead to soft factors which are uncorrected at loop order. In $D>4$ where the IR divergences which affect soft theorems are absent, it has been shown that the sub-leading soft graviton theorem is uncorrected \cite{Sen:2017nim, Laddha:2017ygw}. Loop effects have also been considered in bosonic string theory, also with $D>4$, \cite{Vecchia:2018lek, DiVecchia:2019kle} and in particular it was shown that the dilaton soft theorem is uncorrected by loop effects at any order.

\subsection{Multi-soft Limits}
The emission of $m$ outgoing soft photons with momenta $q_1, \dots, q_m$ is, to leading order in the soft momenta, again given as a product of the amplitude for the hard particles times a  pre-factor, often called the eikonal factor, which is a product of the leading soft terms
\begin{equation}
\mathcal{A}(p,q_1,\dots, q_m) \approx \prod_{j=1}^m \Big[\sum_{i=1}^n Q_i \eta_i \frac{{\epsilon}(q_j)\cdot p_i}{q_j \cdot p_i} \Big] \mathcal{A}(p)~.
\end{equation} 
This eikonal pre-factor can be derived by treating the hard particles as Wilson lines placed along their trajectories, a method which can be usefully generalised to QCD \cite{Korchemskaya:1994qp, Korchemskaya:1996je}. In this limit, the diagrams for soft-photon emission can be shown to exponentiate
\cite{Yennie:1961ad} which implies that low-order perturbative calculations contain all-order information which is important, particularly in the generalisation to the non-Abelian case \cite{Sterman:1981jc, Gatheral:1983cz, Frenkel:1984pz}, \cite{Gardi:2010rn, Mitov:2010rp, Gardi:2013ita}, in phenomenological applications. Exponentiation in the next-to-eikonal approximation, where sub-leading terms in the soft expansion are retained, has been studied for gauge theory, e.g. \cite{Laenen:2008gt, Laenen:2010uz}, and for perturbative gravity \cite{White:2011yy, White:2014qia}. For a thorough discussion of the exponentiation of IR divergences, and further references, see  the recent review \cite{Agarwal:2021ais}.
 
The behaviour of amplitudes under multiple emission of real, soft gluons and gravitons has also been directly studied using on-shell and string theory methods. 
The  tree-level double-soft limits for gluons through sub-leading order were computed using BCFW recursion relations and four-dimensional spinor helicity in \cite{Klose:2015xoa, Volovich:2015yoa} and reproduced in \cite{Georgiou:2015jfa} using the CSW method \cite{Cachazo:2004kj}. Double-soft limits for positive helicity gluons can be given an interpretation as a two-dimensional current-current OPE \cite{He:2015zea}, while multi-soft limits can be interpreted as the OPE of a Sugawara stress tensor and used to derive an analogue of the Knizhnik-Zamolodchikov equation for  amplitudes \cite{McLoughlin:2016uwa}, see also \cite{Fan:2020xjj, Banerjee:2020vnt}. The four-dimensional double soft-limit of gravitons was similarly analyzed in \cite{Klose:2015xoa} through sub-sub-leading order. In general dimensions, the double-soft theorem for gravitons was studied  using the CHY formalism in \cite{Saha:2016kjr} and the sub-leading multiple soft theorem for arbitrary number of soft-gravitons was similarly computed using CHY in \cite{Saha:2017yqi, Chakrabarti:2017ltl} and using Feynman diagrammatics in \cite{Chakrabarti:2017ltl, AtulBhatkar:2018kfi}. Double soft-limits have also been studied in open string theory \cite{ Volovich:2015yoa, DiVecchia:2015bfa} and for massless closed bosonic strings \cite{Marotta:2020oob} where they were shown to be independent of the string $\alpha'$ corrections through sub-leading order. One-loop corrections to the double-soft limit in QCD, formulated as a loop corrected current, were computed in \cite{Zhu:2020ftr, Catani:2021kcy}.
 	
 In the double-soft limit the order in which particles are taken soft can affect the result. Given an amplitude with two soft particles having momenta which scale as $\delta_1 q_1$ and $\delta_2 q_2$ we can study the double-soft limit by expanding in the parameters $\delta_1$ and $\delta_2$. The consecutive soft-limit is defined by first taking $\delta_1 \to 0$ and then $\delta_2 \to 0$, or vice versa, and consequently one can study the symmetrized or anti-symmetrized limit. The simultaneous double-soft limit corresponds to setting $\delta_1=\delta_2=\delta$ and then expanding in~$\delta$. Consecutive double-soft limits are given as ordered products of single soft factors but simultaneous soft limits, particularly sub-leading terms involving particles with different helicities, have correction or ``contact" terms. This issue has arisen when using multi-soft limits to study the algebra of asymptotic symmetries.  For example, anti-symmetrized consecutive double-soft limits of gravitons have been related to the BMS algebra \cite{Distler:2018rwu, Li:2017fsb, Anupam:2018vyu}, while  collinear limits have been used to directly compute the algebra in an ordering free approach \cite{Guevara:2021abz}.

One motivation for the study of the double-soft limit of gauge bosons and gravitons is the similar analysis in the case of theories of spontaneous symmetry breaking, where the double-soft limit of two Goldstone bosons has a non-vanishing limit \cite{Arkani-Hamed:2008owk}. This is related to another important example of a soft theorem, the so-called Adler zero \cite{adler1965consistency} which corresponds to the vanishing of the soft pion amplitude as a consequence of spontaneously broken chiral symmetry. This topic has seen renewed interest and has been actively pursued in recent years giving rise to significant progress in understanding the space of consistent effective field theories by imposing soft limits on tree-level amplitudes, a programme referred to as the soft-bootstrap, see for example \cite{Cheung:2014dqa, Cheung:2015ota, Cheung:2016drk, Elvang:2018dco, Low:2019ynd, Rodina:2021isd}. 

An interesting recent application of multi-soft limits is to the study of classical radiation. The gravitational waveform at a detector far from a radiation generating scattering process can be extracted at late retarded times from the graviton soft theorem \cite{Laddha:2018vbn, Sahoo:2018lxl, Saha:2019tub, Sahoo:2020ryf, Sahoo:2021ctw}. The leading term in the expansion of the waveform corresponds to a permanent change in the metric which is related to the gravitational memory effect \cite{zel1974radiation, christodoulou1991nonlinear} whose relation to soft theorems and asymptotic symmetries was studied in \cite{Strominger:2014pwa}.

\section{Soft theorems and Asymptotic Symmetries}
\label{sec:st-as}

\begin{figure}
	\centering
\includegraphics[scale=1.15]{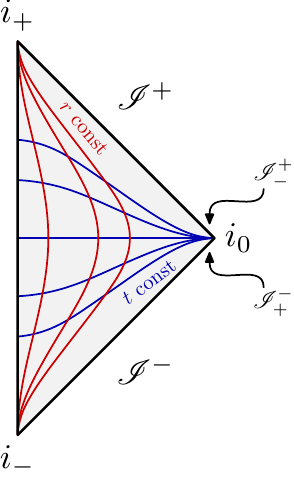}
	\caption{Penrose diagram for Minkowski spacetime.}
	\label{fig:Pen_flat}
\end{figure}

The universal properties of scattering amplitudes discussed in section  \ref{sec:soft} arise naturally from the conservation laws associated with asymptotic or large gauge symmetries. In this section we give a brief account of the fundamental principles underlying this remarkable equivalence. We refer the reader to \cite{Strominger:2017zoo} for a pedagogical review.

The analysis of asymptotic symmetries in asymptotically flat gravity has a long history dating back to the works of Bondi, van der Burg, Metzner and Sachs \cite{Bondi:1962px, Sachs:1962wk, Sachs:1962zza}. At spatial infinity, conserved quantities such as mass and angular momentum can be defined in the Hamiltonian formalism \cite{Arnowitt:1962hi}. They are in one-to-one correspondence with asymptotic symmetries, namely the set of diffeomorphisms that preserve specified boundary conditions and act non-trivially on the phase space or physical data of the theory. At null infinity ($\mathscr{I}$), this analysis is complicated by the fact that charges can be radiated away. Lee, Iyer, Wald and Zoupas showed that conserved quantities which live on cross-sections of null infinity and generate the symmetry can nevertheless be constructed under certain conditions  \cite{Lee:1990nz,Iyer:1994ys,Wald:1999wa}.\footnote{~This approach can be generalised to gauge theories in asymptotically flat backgrounds where the importance of large gauge transformations was only more recently recognized \cite{He:2014cra}. } In practice, the analysis is often subtle, relies on a choice of asymptotic falloffs and gauge choices and remains an open problem in the full interacting theory.

In the following, we will assume that the charges associated with the asymptotic symmetries at future or past null infinity ($\mathscr{I}^+$ or $\mathscr{I}^-$) have been constructed via a canonical analysis or otherwise. These charges are defined on cross-sections of $\mathscr{I}$ taken to be the past and future boundaries $\mathscr{I}^+_-$ and $\mathscr{I}^-_+$ of, respectively, $\mathscr{I}^+$ and $\mathscr{I}^-$. We denote them by $Q^+$ and $Q^-$ and suppress their dependence on the asymptotic data and variation of the data under asymptotic symmetries.
Charge conservation is the statement that
\begin{equation}
\label{cpm}
Q^+ = Q^-.
\end{equation}
This conservation law is far from obvious in general, but it was remarkably shown to be implied by a set of matching conditions \cite{Strominger:2013jfa} obeyed in a wide range of physical examples \cite{Strominger:2013lka,He:2014laa,Kapec:2014opa,Lysov:2014csa,He:2014cra,He:2015zea,Kapec:2015ena,Dumitrescu:2015fej,Henneaux:2018hdj,Prabhu:2018gzs,Prabhu:2019fsp,Henneaux:2019yqq,Satishchandran:2019pyc,Prabhu:2021cgk}. Imposing that the $\mathcal{S}$-matrix obeys the symmetry \eqref{cpm} then leads to 
\begin{equation}
\label{cons-charges}
\langle  {\rm out} | Q^+ \mathcal{S} - \mathcal{S} Q^- | {\rm in} \rangle = 0.
\end{equation}
Provided the geometry reverts to the vacuum at the future and past boundaries of $\mathscr{I}^+$ and $\mathscr{I}^-$ respectively, the charges can be split into a ``soft'' part which is linear in the asymptotic data, and a nonlinear ``hard'' part 
\begin{equation}
\label{soft-hard-split}
Q^{\pm} = Q^{\pm}_S + Q^{\pm}_H.
\end{equation}
Substituting \eqref{soft-hard-split} 
into \eqref{cons-charges}, one finds
\begin{equation}
\label{cc-soft}
\langle  {\rm out} | Q^+_S \mathcal{S} - \mathcal{S} Q^-_S | {\rm in} \rangle = - \langle  {\rm out} | Q^+_H \mathcal{S} - \mathcal{S} Q^-_H| {\rm in} \rangle.
\end{equation}
Upon quantization, the left hand side of \eqref{cc-soft} can be identified with a particular soft mode of the gauge or gravitational field, while the right hand side reproduces the corresponding soft factor \cite{He:2014laa}. 
As such, the validity of soft theorems a posteriori justifies the conservation law \eqref{cpm}. 

Soft theorems have revealed richer than expected asymptotic/large gauge symmetry structures. In gravity, generalizations of the extended BMS group \cite{Barnich:2009se,Barnich:2010ojg,Barnich:2011mi} have been studied at null \cite{Campiglia:2014yka, Flanagan:2015pxa,Conde:2016rom,Campiglia:2016efb, Bakhoda:2019htc,Compere:2019gft, Compere:2018ylh, Campiglia:2020qvc, Freidel:2021fxf, Freidel:2021cjp, Campiglia:2021bap, Freidel:2021dfs}, spacelike \cite{Troessaert:2017jcm, Henneaux:2018cst, Henneaux:2018hdj, Henneaux:2018mgn, Henneaux:2019yax,Prabhu:2019daz, Brocki:2021ieh} and timelike \cite{Campiglia:2015lxa,Campiglia:2015kxa, AH:2020rfq, Chakraborty:2021sbc} infinity. Gravitational analogs of electric-magnetic duality have been proposed in \cite{Godazgar:2019dkh, Godazgar:2019ikr, Choi:2019sjs, Bunster:2019mup, Huang:2019cja, Godazgar:2020gqd, Godazgar:2020kqd, Oliveri:2020xls, Kol:2020zth}. Observable signatures  of these symmetries include memory effects at null infinity \cite{Strominger:2014pwa,Pasterski:2015tva,Pate:2017fgt,Hamada:2018cjj, Mao:2018xcw, Compere:2019odm, Satishchandran:2019pyc} or near black hole horizons \cite{Setare:2016msj,Donnay:2018ckb, Grumiller:2019ygj, Choi:2019fuq, Rahman:2019bmk, Bhattacharjee:2020vfb, Giribet:2020rvd}, as well as in shockwave backgrounds with accelerated observers  \cite{OLoughlin:2018ebk, Compere:2019rof, Bhattacharjee:2019jaf, Ferreira:2020whz, Gray:2021dfk}.  Counterparts of asymptotic symmetries and memory effects in higher-dimensional asymptotically flat spacetimes have been analyzed in \cite{Pate:2017fgt, Marotta:2019cip, Campoleoni:2020ejn,Fuentealba:2021yvo}. Connections to the symmetry groups of negatively  and positively curved spacetimes, as well as the gauge-gravity duality have been explored in \cite{Poole:2018koa, Ball:2019atb, Compere:2020lrt, Lowe:2020qan, Banerjee:2020dww, Banerjee:2021llh}, while implications for the black hole information paradox have been discussed in \cite{Hawking:2016msc, Strominger:2017aeh, Haco:2018ske,Laddha:2020kvp, Raju:2020smc, Raju:2021lwh}. 

In gauge theory, the asymptotic symmetry structure and the matching conditions have been studied at null \cite{He:2014cra,Kapec:2015ena,Freidel:2019ohg}, spacelike \cite{Esmaeili:2019hom, Henneaux:2019yqq} and timelike \cite{Campiglia:2015kxa} infinity. Magnetic symmetries and the electric-magnetic duality are discussed in \cite{Strominger:2015bla,Freidel:2018fsk,Huang:2019cja,Henneaux:2020nxi,Geiller:2021gdk}.  The symmetries of higher-dimensional gauge theories and their relation to soft theorems have been worked out in \cite{He:2019jjk, Henneaux:2019yqq}. The relation between asymptotic symmetries and loop corrections to the sub-leading soft photon theorem has been addressed in \cite{Campiglia:2019wxe, Banerjee:2021llh, AtulBhatkar:2020hqz}, while multi-soft limits were recovered from nested charge commutators in \cite{Distler:2018rwu,Liu:2021dyq}. The relation to gauge theories in AdS in the flat space limit has been studied in \cite{Esmaeili:2019mbw,Hijano:2020szl}. Memory effects to all orders in a large-$r$ expansion were discussed in \cite{Mao:2021eor}.

 In the remainder of this section we outline some of the simplest examples of asymptotic symmetry enhancements in (massless) QED and gravity. The discussion is meant to motivate the construction of celestial amplitudes which will be the focus of the following sections.

\subsection{QED}

Consider Maxwell theory in Minkowski space 
\begin{equation}
\label{MS}
ds^2 = - du^2 -2 du dr + r^2 \gamma_{AB} dx^A dx^B,
    \end{equation}
where $x^A$ are coordinates and $\gamma_{AB}$ is the unit metric on the sphere and $u$ is the retarded time. The leading soft photon theorem is associated with angle-dependent large gauge transformations \cite{He:2014cra}
\begin{equation} 
\label{lgt}
A_{B}^{(0)} \rightarrow A_{B}^{(0)} + \partial_{B} \varepsilon(x^A),
\end{equation}
where $A_{B}^{(0)}$  
is the leading $\calO(r^0)$ component in a large-$r$ expansion of the transverse/radiative gauge potential. The charges generating \eqref{lgt} are 
\begin{equation} 
\label{ec}
Q^{+}_{\varepsilon} = \frac{1}{e^2}\int_{\mathscr{I}^{+}_{-}} d^2x \sqrt{{\rm det} (\gamma)} \,\varepsilon \,F_{ru}^{(2)},
\end{equation}
where $F_{ru}^{(2)}$ is the $\mathcal{O}(r^{-2})$ component of $F_{ru}$ in a large-$r$ expansion. 
For constant $\varepsilon$, the large gauge charges \eqref{ec} reduce to the standard electric charges. A similar argument yields the large gauge charges $Q^-$ on $\mathscr{I}^-$ where $u$ is replaced by the advanced time $v$. 
Using the constraint equation 
\begin{equation}
\partial_u F_{ur}^{(2)} = D^A F_{uA}^{(0)} + e^2 j_u^{(2)}
\end{equation}
on $\mathscr{I}^+$ (and its analog on $\mathscr{I}^-),$ the charges $Q^{+}$ ($Q^-$) split into a soft piece linear in $F_{uA}^{(0)}$ ($F_{vA}^{(0)}$) and a hard piece involving the current $j_u^{(2)}$ ($j_v^{(2)}$) (quadratic in the matter fields). Here $D_A$ denotes the covariant derivative on the sphere which reduces to $\partial_A$ on the plane.  For a particular choice of $\varepsilon(x^A)$, upon quantization and the use of crossing symmetry $\langle {\rm out}|a_\pm (q)\mathcal S|{\rm in}\rangle =\langle {\rm out}|\mathcal S a_\mp^\dagger(- q)|{\rm in}\rangle$, the conservation law~\eqref{cc-soft} associated with~\eqref{ec} can be recast as the leading soft photon theorem \cite{He:2014cra}
\begin{equation}
\label{eq:sphth}
    \lim_{\delta \rightarrow 0} \delta \langle {\rm out}|a_{\pm}(\delta q) \mathcal{S} |{\rm in}\rangle = S^{(0)}_{\pm} \langle {\rm out}| \mathcal{S} |{\rm in} \rangle,
\end{equation}
where $S^{(0)}_{\pm}$ is given in \eqref{eq:soft-photon-factors}. Conversely, \eqref{eq:sphth} leads to the conservation law \eqref{cc-soft} upon picking out particular modes on the sphere via integration against $\partial_{B}\varepsilon$ depending on the helicity of the soft photon insertion.

\subsection{Gravity}

A generic four-dimensional asymptotically flat spacetime is described in retarded Bondi coordinates by the metric \cite{Barnich:2009se,Barnich:2010eb}
\begin{equation}
\label{AFS}
ds^2 = -2 e^{2\beta} du (dr + \Phi du) + g_{AB} \left(dx^A - \frac{\Upsilon^A}{r^2} du \right)\left(dx^B - \frac{\Upsilon^B}{r^2} du \right).
    \end{equation}
The functions appearing in~\eqref{AFS} have the following large~$r$ expansion 
    \begin{equation}
    \label{mc}
        \begin{split}
\Phi &= F - \frac{m_B}{r} + \mathcal{O}(r^{-2}), \\
\beta &= \frac{b}{r^2} + \mathcal{O}(r^{-3}), \\
\Upsilon^A &= U^A - \frac{2\gamma^{AB}}{3r}\left(N_B+ C_{BC} U^C + \partial_B b \right) + \mathcal{O}(r^{-2}), \\
g_{AB} &= r^2 \gamma_{AB} + r C_{AB} +  \mathcal{O}(r^{0}),
\end{split}
    \end{equation}
where $m_B,b,U^A,N_B,C_{AB}$ are functions of $u$ and $x^A$ while $F$ and $\gamma_{AB}$ depend only on $x^A$ ($F$ is related to the curvature of $\gamma_{AB}$ by Einstein's equations eg. $F=\frac{1}{2}$ (sphere) or $F=0$ (plane)).
The asymptotic symmetries of~\eqref{AFS} are generated by \cite{Barnich:2009se} 
\begin{equation} 
\label{asg}
\left. \xi\right|_{\mathscr{I}^+} = \left(f + \frac{u}{2} D_A Y^A\right)\partial_u + Y^A \partial_{A}
\end{equation}
and analogous vector fields at $\mathscr{I}^-$ with $f$, $Y^A$ antipodally matched functions and vector fields on the sphere respectively~\cite{Strominger:2013jfa}. The diffeomorphism~\eqref{asg} with $Y^A = 0$ generates supertranslations -- an infinite-dimensional enhancement of the Poincar\'e translations.
Provided the boundary conditions do not allow for leading changes in $g_{AB}$, the vector fields $Y^A$ are restricted to be holomorphic/antiholomorphic and~\eqref{asg} with $f = 0$ generate two copies of the Virasoro algebra \cite{Barnich:2011mi}. Otherwise, the former condition is relaxed and they  generate {\rm Diff}($S^2$)~\cite{Campiglia:2014yka}.\footnote{Meromorphic vector fields $Y^A$ also violate the boundary conditions, albeit at isolated points.} 
Both cases represent infinite-dimensional enhancements of Lorentz symmetry. For arbitrary $f,~ Y^A$, the diffeomorphisms~\eqref{asg} generate the generalised BMS algebra \cite{Compere:2018ylh}.\footnote{~A further extension allowing for Weyl rescalings of the transverse metric was proposed in~\cite{Freidel:2021fxf}.} 

The leading soft graviton theorem was shown in~\cite{He:2014laa} to be equivalent to the conservation of the supertranslation charges 
\begin{equation} 
\label{lead-sc}
Q_{f}^{\pm} = \frac{8}{\kappa^2}\int_{\mathscr{I}^{\pm}_{\mp}}d^2x  \sqrt{{\rm det} (\gamma)} f m_B,
\end{equation}
which reduces to the ADM mass for constant $f$.
The integral~\eqref{lead-sc} can be extended over all of $\mathscr{I}^{+}$ using the constraint\footnote{~If gravity is coupled to matter, a matter stress tensor term will appear on the right hand side.}
\begin{equation} 
\partial_u m_B = \frac{1}{4} D_A D_B N^{AB}  - \frac{1}{8} N_{AB} N^{AB},
\end{equation}
and its analog near $\mathscr{I}^-$~\cite{Strominger:2013jfa}.
Then, as described previously, the supertranslation charge~\eqref{lead-sc} splits into a soft part linear in $N_{AB}$ and a hard part quadratic in $N_{AB}$. After quantization and for an appropriate choice of $f$, when substituted into the conservation law~\eqref{cc-soft} it can be shown to reproduce the soft graviton theorem
\begin{equation}
\label{eq:lsgt}
    \lim_{\delta \rightarrow 0} \delta \langle {\rm out}|a_{\pm}(\delta q) \mathcal{S} |{\rm in}\rangle = S^{(0)}_{\pm} \langle {\rm out}| \mathcal{S} |{\rm in} \rangle
\end{equation}
with $S^{(0)}_{\pm}$ given in~\eqref{eq:soft-graviton-factors}. 
Similarly, invoking the constraint equation for $\partial_u N_A$ 
the conservation law associated with the superrotation charge 
\begin{equation} 
\label{sub-sc}
Q_{Y}^{\pm} =  \frac{4}{\kappa^2}\int_{\mathscr{I}^{\pm}_{\mp}} d^2x \sqrt{{\rm det} (\gamma)} Y^A N_A,
\end{equation}
 was shown to be implied by the sub-leading soft graviton theorem \cite{Kapec:2014opa} 
\begin{equation}
\label{eq:slsgth}
    \lim_{\delta \rightarrow 0} (1 + \delta \partial_{\delta}) \langle {\rm out}|a_{\pm}(\delta q) \mathcal{S} |{\rm in}\rangle = S^{(1)}_{\pm} \langle {\rm out}| \mathcal{S} |{\rm in} \rangle
\end{equation}
with $S^{(1)}_{\pm}$ given in \eqref{eq:soft-graviton-factors}. This has remarkable implications as we now discuss.
\setcounter{footnote}{0} 

In fact, the soft component of the charge \eqref{sub-sc} can be related to the generator of a two-dimensional local conformal symmetry on the sphere by choosing a meromorphic vector field $Y^A$, while the soft theorem~\eqref{eq:slsgth} can be recast as a Ward identity of that symmetry. Let us focus on scattering of massless particles with momenta 
\begin{equation}\label{eq:pi}
    p_i^\mu= \eta_i \omega_i (1+z_i\bz_i,z_i+\bz_i,i(\bz_i-z_i),1-z_i\bz_i), 
\end{equation}
where henceforth we absorb the $\eta_i=\pm 1$ into the definition of the momenta.
We denote the amplitude without the soft graviton by  $\langle\prod_{i=1}^n \mathcal{O}^\pm_{\ell_i}(\omega_i, z_i, \bar{z}_i)\rangle$ to make the dependence on the asymptotic on-shell particle states of energy $\omega_i$ and helicity $\ell_i$ piercing the celestial sphere at a point $(z_i, \bar{z}_i)$ at $\mathscr{I}^\pm$ as well as the $\eta_i=\pm$ label manifest.  For simplicity, here and henceforth, we will use flat coordinates in~\eqref{AFS} with $F = 0$.
Integrating the sub-leading soft theorem~\eqref{eq:slsgth} against an appropriate function on the sphere\footnote{~This operation coincides with the shadow transform of a conformally soft graviton~\cite{Pasterski:2017kqt,Donnay:2018neh}.}
one finds it can be put into the form~\cite{Kapec:2016jld}
\begin{equation}
\label{eq:Tzz-ward}
     \langle T_{zz} \prod_{i=1}^n \mathcal{O}^\pm_{\ell_i}(\omega_i, z_i, \bar{z}_i)\rangle = \sum_{k = 1}^n\left[\frac{\hat{h}_k}{(z - z_k)^2} + \frac{\partial_{z_k}}{z - z_k} \right]\langle \prod_{i=1}^n \mathcal{O}^\pm_{\ell_i}(\omega_i, z_i, \bar{z}_i) \rangle.
\end{equation}
Here $T_{zz}$ denotes the insertion of the soft graviton mode picked out by the operator $1+\delta \partial_\delta$ (and smeared over the sphere) and we have introduced
\begin{equation} 
\label{eq:weight-ops}
\hat{h}_i = \frac{1}{2}\left(-\omega_i \partial_{\omega_i} + \ell_i\right).
\end{equation}

The result \eqref{eq:Tzz-ward} is strikingly similar to the Ward identity of the stress tensor in a two-dimensional conformal field theory (CFT). On the one hand this is expected since the four-dimensional Lorentz group acts as the global conformal group on the two-dimensional celestial sphere at null infinity 
\begin{equation}
    {\rm SO}^+(1,3)\simeq {\rm SL}(2,\mathbb{C})/\mathbb{Z}_2.
\end{equation}
On the other hand, we see that the sub-leading soft graviton theorem encodes the enhancement of Lorentz symmetry to Virasoro or Diff($S^2$). This is analogous to the enhancement of conformal symmetry to Virasoro in two-dimensional CFT and hints at a holographic principle for asymptotically flat spacetimes. The first step in making this equivalence more precise is to recast the scattering matrix into a basis that diagonalizes the weights \eqref{eq:weight-ops}. 
As we will see, this can be achieved by switching from the standard basis of asymptotic momentum eigenstates to a basis of boost eigenstates.

\section{Celestial Amplitudes}
\label{sec:celestial-amplitudes}

A profusion of fascinating results in scattering amplitudes has been obtained using on-shell methods in a translation invariant basis. In this section we review novel insights that arise by making 4D Lorentz symmetry or, equivalently, 2D global conformal symmetry on the celestial sphere manifest. The resulting {\it celestial amplitudes} exhibit a rich symmetry structure and behave in many ways as correlation functions in a conformal field theory - albeit one of a more exotic type referred to as ``celestial conformal field theory" (CCFT). 
This paves a path towards a holographic principle for asymptotically flat spacetimes known as {\it celestial holography}.

\subsection{The Holographic Map}
\label{subsec:holomap}

The holographic flavor of the $\calS$-matrix becomes manifest when the asymptotic states are taken to be boost eigenstates 
\begin{equation}\label{eq:4d2dmap}
   {}_{\rm boost}\langle {\rm out} | \calS |{\rm in}\rangle_{\rm boost}=\langle \calO^\pm_{\Delta_1,J_1}(z_1,\bz_1)\cdots \calO^\pm_{\Delta_n,J_n}(z_n,\bz_n)\rangle_{\rm CCFT}.
\end{equation}
This change of basis recasts it as a correlation function of operators $\calO^{\pm}_{\Delta_i,J_i}$ inserted at points $(z_i,\bz_i)$ on the celestial sphere that are labeled by SL(2,$\mathbb{C}$) conformal dimensions~$\Delta_i$ and spins~$J_i$ and a sign distinguishing between out~(+) and in~(-) states. Starting from momentum-space amplitudes, celestial amplitudes are constructed as follows.

 A massless momentum  in Minkowski space can be written as $ p^\mu=\pm \omega q^\mu(z,\bz)$
with an energy scale $\omega$,  a sign for outgoing (+) versus incoming (-), and a null vector directed at a point $(z,\bz)$ on the celestial sphere
\begin{equation}\label{eq:masslessmom}
q^\mu(z,\bz)= (1+z\bz,z+\bz,i(\bz-z),1-z\bz).
\end{equation}
The celestial amplitude~\eqref{eq:4d2dmap} for a scattering process of $n$~massless particles with spins~$s_i$ and momenta $p^\mu_i=\pm \omega_i q^\mu_i(z_i,\bz_i)$ is simply a Mellin transform with respect to the energies of each external scattering state \cite{Pasterski:2016qvg, Pasterski:2017kqt} 
\begin{equation}\label{eq:4d2dmapmassless}
  \tcalA_n(\Delta_i,J_i;z_i,\bz_i)\equiv \prod_{k=1}^n \int_0^\infty \frac{d\omega_k}{\omega_k}\omega_k^{\Delta_k} \calA_n(\omega_i, \ell_i,z_i,\bz_i)
\end{equation}
of the momentum-space amplitude $\calA_n(\omega_i,\ell_i, z_i, \bar{z}_i)$ which includes the momentum conserving delta function $\delta^{(4)}(\sum_{i=1}^n p_i^\mu)$. 
This map trades energies $\omega_i$ for boost weights $\Delta_i$ (or Rindler energies), while the 4D helicities~$\ell_i=\pm s_i$ become 2D spins~$J_i$.

Celestial amplitudes for massive particles are more involved.
The mass-shell of a massive momentum is parameterized by a hyperbolic slice $H_3$ of Minkowski space. An on-shell momentum $  p^\mu=\pm m\hat p^\mu(y,w,\bw)$
with $\hat p^2=-1$ is embedded into the upper branch ($\hat p^0>1$) of the unit hyperboloid in Minkowski space as
\begin{equation}\label{eq:hatp}
    \hat p^\mu(y,w,\bw)=\frac{1}{2y}(1+y^2+w \bw,w+\bw,i(\bw-w),1-y^2-w \bw).
\end{equation}
The celestial amplitude~\eqref{eq:4d2dmap} for an $n$-point scattering amplitude of massive scalars  with momenta $p^\mu_i=\pm m_i \hat p^\mu_i(y_i,w_i,\bw_i)$ is realized by the integral transform
\begin{equation}\label{eq:4d2dmapmassive}
  \tcalA_n(\Delta_i;z_i,\bz_i)\equiv \prod_{k=1}^n \int_{{H}_3} \frac{d^3\hat p_k}{\hat p_k^0} G_{\Delta_k}(\hat p_k(y_k,w_k,\bw_k);q_k(z_k,\bz_k)) \calA_n(m_i \hat p_i)\,,
\end{equation}
where $ G_{\Delta}(\hat p(y,w,\bw);q(z,\bz))=(-q\cdot \hat p)^{-\Delta}$
is the scalar bulk-to-boundary propagator in $H_3$~\cite{Costa:2014kfa}. For generic massive spin-$s$ scattering the 
the analogue of~\eqref{eq:4d2dmapmassive} has been worked out in \cite{Pasterski:2016qvg,Law:2020tsg,Narayanan:2020amh,Iacobacci:2020por}.

While momentum-space amplitudes describe probabilities for scattering plane waves and exhibit manifest translation symmetry, celestial amplitudes
scatter wavefunctions that transform with definite $(\Delta,J)$ under an SL$(2,\mathbb{C})$ Lorentz transformation. In what follows we review the construction of the so-called {\it conformal primary wavefunctions} and how they give rise to the boundary operators $\calO^{\pm}_{\Delta,J}$ in~\eqref{eq:4d2dmap} before discussing some universal properties of celestial amplitudes. We restrict to the scattering of massless particles henceforth.

\subsection{Conformal Primary Wavefunctions and Operators}
\label{subsec:cpw}

The celestial operators appearing in~\eqref{eq:4d2dmap} 
are defined via the map
\begin{equation}
\label{eq:2Dop}
\mathcal{O}^{\pm}_{\Delta,J}(z,\bz)=i(\hat{O}^s(X^\mu),\Phi^{s}_{\Delta,J}(X_\mp^\mu;z,\bz)^*)_{\Sigma}.
\end{equation}
Here $\hat O^s(X^\mu)$ is a spin-$s$ bulk operator that creates a single particle state when acting on the vacuum. The inner product  $(\cdot \,,\cdot)_\Sigma$ is defined on a Cauchy slice $\Sigma$ and can be constructed from the symplectic product for spin-$s$ wavefunctions via $\Omega(\Phi,\Phi')=i(\Phi,{\Phi'}^*)_\Sigma$ where $*$ denotes complex conjugation. For scalar wavefunctions it corresponds to the standard Klein-Gordon inner product\footnote{~An alternate inner product involving the shadow transform was recently advocated for in~\cite{Crawley:2021ivb} and leads to a reorganization of the celestial CFT data. Its implications are an open problem.}, while it is appropriately generalised for arbitrary spin~\cite{Pasterski:2017kqt,Law:2020tsg,Iacobacci:2020por,Narayanan:2020amh}.
The wavefunctions $\Phi^s_{\Delta,J}(X_\pm^\mu;z,\bz)$ satisfy the linearised spin-$s$ equations of motion and are constructed from Mellin transforms of plane waves
\begin{equation}\label{eq:phi}
     \int_0^\infty d\omega \omega^{\Delta-1} e^{\pm i \omega q\cdot X_\pm}=\frac{(\mp i)^\Delta \Gamma(\Delta)}{(-q\cdot X_\pm)^\Delta}\,.
\end{equation}
The $\pm$ label distinguishes between outgoing and incoming modes, and is selected by the analytic continuation $X^\mu_\pm=X^\mu\pm i\varepsilon\{-1,0,0,0\}$ which serves as a regulator (that we omit henceforth).
The wavefunctions $\Phi^s_{\Delta,J}(X^\mu;z,\bz)$ are called {\it conformal primary wavefunctions} owing to their transformation properties under the appropriate representations of Poincar\'e  in both the bulk ($X^\mu$) and boundary ($z,\bz$) coordinates. Namely, under simultaneous SL(2,$\mathbb{C}$) Lorentz transformations 
 \begin{equation}\label{eq:SL2C}
X^\mu\mapsto \Lambda^\mu_{~\nu}X^\nu\,,\quad z\mapsto \frac{a z+b}{cz+d}\,,\quad \bz\mapsto \frac{{\bar a} \bz+{\bar b}}{{\bar c}\bz+{\bar d}}\,,
 \end{equation}
they transform as two-dimensional conformal primaries
\begin{equation}\label{eq:CPWdef}
\Phi^s_{\Delta,J}\Big(\Lambda^{\mu}_{~\nu} X^\nu;\frac{a z+b}{cz+d},\frac{{\bar a} \bz+{\bar b}}{{\bar c}\bz+{\bar d}}\Big)=(cz+d)^{\Delta+J}({\bar c}\bz+{\bar d})^{\Delta-J}D_s(\Lambda)\Phi^{ s}_{\Delta,J}(X^\mu;z,\bz)\,,
\end{equation}
with conformal dimension~$\Delta$ and spin~$J$.
Here $\{a,b,c,d\}\in \mathbb{C}$ with $ad-bc=1$, and $D_s(\Lambda)$ is the spin-$s$ representation of the Lorentz algebra.

Conformal primary wavefunctions are constructed from~\eqref{eq:phi} dressed by the appropriate frame fields. 
We consider scalar conformal primary wavefunctions
\begin{equation}\label{eq:varphi}
    \varphi_\Delta=\frac{1}{(-q\cdot X)^\Delta}\,,
\end{equation}
where compared to~\eqref{eq:phi} we dropped the normalisation factor and omitted the $\pm$~label. 
Natural polarisation vectors constructed from~\eqref{eq:masslessmom} are $\sqrt{2}\epsilon_+^\mu=\partial_z q^\mu$ and $\sqrt{2}\epsilon_-^\mu=\partial_{\bar z} q^\mu$ but these do not have definite SL(2,$\mathbb{C}$) weights. Instead we construct spacetime dependent polarisation vectors as~\cite{Pasterski:2020pdk} 
\begin{equation}\label{eq:mmbar}
m^\mu=\epsilon^\mu_++\frac{\epsilon_+\cdot X}{(-q\cdot X)} q^\mu\,, ~~~\bar{m}^\mu=\epsilon^\mu_- +\frac{\epsilon_-\cdot X}{(-q\cdot X)} q^\mu\,,
\end{equation}
which satisfy $m\cdot\bar{m}=1$ and transform with $\Delta=0$ and, respectively, $J=+1$ and $J=-1$.\footnote{~One can further complete the vectors~\eqref{eq:mmbar} into a null tetrad $\{l,n,m,\bm\}$ for Minkowski space by~\cite{Pasterski:2020pdk}
\begin{equation}\label{eq:tetrad}
l^\mu=\frac{q^\mu}{-q\cdot X}\,, ~~~n^\mu=X^\mu+\frac{X^2}{2}l^\mu\,, 
\end{equation}
which satisfy $l\cdot n=-1$ and transform with $\Delta=J=0$.}
\setcounter{footnote}{1}
We define spin-one and spin-two conformal primary wavefunctions by
\begin{equation}\begin{array}{ll}\label{eq:CPWs}
     A_{\Delta,J=+ 1}=m \varphi_{\Delta} \,,&\quad A_{\Delta,J=- 1}=\bar{m} \varphi_{\Delta}\,, \\
    h_{\Delta,J=+2}=m m \varphi_{\Delta} \,,&\quad  h_{\Delta,J=-2}=\bar{m} \bar{m} \varphi_{\Delta}\,.
\end{array}
\end{equation}
An analogous construction for conformal primary wavefunctions with half-integer spin using a decomposition of the null tetrad~\eqref{eq:mmbar}-\eqref{eq:tetrad} into a spin frame can be found in~\cite{Pasterski:2020pdk}. 

A priori conformal primary wavefunctions have arbitrary complex dimension $\Delta$.
When the conformal dimension lies on the principal continuous series $\Delta\in 1+i\mathbb{R}$ of the SL(2,$\mathbb{C}$) Lorentz group they have been shown to form a complete $\delta$-function normalisable basis~\cite{Pasterski:2017kqt}.  
Another complete basis of conformal primary wavefunctions for the same range of conformal dimension is obtained from the shadow transform of~\eqref{eq:phi} and~\eqref{eq:CPWs} which acts on the corresponding celestial operators as
\begin{equation}\label{2dShadowTransform}
\widetilde{\mathcal O}_{\tDelta,\tJ}(z,\bz)=\frac{k_{\Delta,J}}{2\pi} \int d^2z' \frac{\mathcal O_{\Delta,J}(z',\bz')}{(z-z')^{2-\Delta-J}(\bz-\bz')^{2-\Delta+J}}\,
\end{equation}
where $\tDelta=2-\Delta$, $\tJ=-J$,
and squares to $(-1)^{2J}$ for suitable normalisation $k_{\Delta,J}$. 
Conformal primaries with arbitrary $\Delta\in \mathbb{C}$ can be obtained via contour integrals on the principal series~\cite{Donnay:2020guq}. Of particular importance for the discussion of symmetries in celestial CFT are conformal primaries with conformal dimensions in the (half-)integers as we will see in section~\ref{sec:Celestial-symmetries}.

We conclude this section with a discussion of special properties of conformal primary wavefunctions. 
First note that the wavefunctions~\eqref{eq:CPWs} with~\eqref{eq:varphi} satisfy a Kerr-Schild double copy~\cite{Pasterski:2020pdk}.
The metric
\begin{equation}\label{eq:gKerrSchild}
    g_{\Delta,J;\mu\nu}=\eta_{\mu\nu}+h_{\Delta,J;\mu\nu}
\end{equation}
is of Kerr-Schild form~\cite{Kerr2009}, i.e. it can be written as the Minkowski metric $\eta_{\mu\nu}$ plus a function ($\varphi_\Delta$) satisfying the scalar wave equation dressed by two copies of a so-called Kerr-Schild vector ($m$ or $\bar m$) which has the property that it is null and geodesic with respect to both the Minkowski and the full metric. 
The Kerr-Schild double copy relates a class of Kerr-Schild spacetimes to solutions of Maxwell's equation and is a powerful tool for identifying exact solutions of Einstein's equations~\cite{Monteiro:2014cda}. 
Exploiting the famous property that Kerr-Schild metrics linearize the Ricci tensor with mixed  indices, the Ricci tensor for~\eqref{eq:gKerrSchild} can be shown to vanish $ R_{\Delta,J;\mu\nu}=0$. The spin-two primary $h_{\Delta,J=\pm2}$ thus gives rise to an exact solution to the vacuum Einstein equations.

Another interesting property of the wavefunctions~\eqref{eq:CPWs} is that they exhibit definite (anti-)self duality~\cite{Donnay:2018neh}. They thus naturally satisfy a Weyl double copy~\cite{Pasterski:2020pdk} which relates the (anti)self-dual part of the curvature of the spacetime to the (anti)self-dual part of an electromagnetic field strength~\cite{Luna:2018dpt}. The Weyl and Kerr-Schild double copy are reviewed in the SAGEX review Chapter~14 \cite{Kosower:2022yvp}. See also~\cite{Adamo:2021dfg,Godazgar:2021iae} for the classical double copy at null infinity. 

Building on the above results and relaxing the conditions for {\it radiative} conformal primary wavefunctions which have $|J|=s$, one finds that~\eqref{eq:gKerrSchild} with $h_{\Delta,J}$ given by {\it generalised} conformal primaries satisfying~\eqref{eq:CPWdef} with $|J|<s$ describe known exact solutions to Einstein's equations such as ultraboosted black holes and shockwaves~\cite{Pasterski:2020pdk}. 

\subsection{Analytic Structure of Celestial Amplitudes}
\label{sec:analytic-structure}

The prescription described in previous sections in principle allows for any $\mathcal{S}$-matrix to be expressed in a conformal primary basis and hence for arbitrary $n$-point momentum-space scattering amplitudes to be mapped to celestial amplitudes. 
In this section we review the case of generic massless four-point scattering. 
We focus on scalar scattering for simplicity, while the generalization to the spinning case follows upon dressing the amplitudes with the appropriate frame fields as described in section~\ref{subsec:cpw}. 
We will see that all information about the scattering is encoded in the analytic structure of celestial amplitudes as functions of a net boost weight $\beta$ dual to the center of mass energy and a conformally invariant cross-ratio on the sphere $r$ 
related to the bulk scattering angle.

A generic four-point momentum-space scattering amplitude is a function of the Mandelstam invariants $s, t, u$ defined in terms of the particle momenta $p_i$ as 
\begin{equation}
\begin{split}
s = -(p_1 + p_2)^2, \quad
t = -(p_1 + p_3)^2, \quad
u = -(p_1 + p_4)^2.
\end{split}
\end{equation}
Momentum conservation additionally implies $s + t + u = 0$ so that any massless four-point scattering amplitude $\mathcal{A}_4$ can be written as 
\begin{equation}
\label{eq:mssa}
\mathcal{A}_4(p_1, p_2, p_3, p_4) = \mathcal{M}(s, t) \delta^{(4)}\left(p_1 + p_2 + p_3 + p_4 \right).
\end{equation}
Celestial amplitudes are obtained from \eqref{eq:mssa} upon parameterizing the momenta as in \eqref{eq:masslessmom} and evaluating the Mellin transforms with respect to the external energies. It will be convenient to define 
\begin{equation}
\beta \equiv \sum_{i = 1}^4 \Delta_i, \quad r \equiv -\frac{t}{s} = \frac{z_{13} z_{24}}{z_{12} z_{34}}.
\end{equation}
Momentum conservation implies that the resulting celestial amplitude is distributional with non-trivial support on $r = \bar{r}$ and allows for three of the four Mellin integrals to be easily evaluated. The final result takes the simple form 
\begin{equation} 
\label{eq:gca}
\widetilde{\mathcal{A}}_4(\Delta_i, z_i, \bar{z}_i) = K(\Delta_i, z_i, \bar{z}_i) \int_0^{\infty} d\omega \omega^{\beta - 1} \mathcal{M}( \omega^2, -\omega^2 r)\delta(r - \bar{r}).
\end{equation}
Here $K(\Delta_i, z_i, \bar{z}_i)$ includes a conformally covariant structure, as well as a universal conformally invariant function. Its explicit form can be found in~\cite{Schreiber:2017jsr,Arkani-Hamed:2020gyp}.

Since~\eqref{eq:gca} involves an integral over all center of mass energies, we see that in order for the celestial amplitudes to be well defined, the momentum-space amplitudes need to have sufficiently soft behaviour at high energies. In the remainder of this section we discuss how the UV behaviour of momentum-space scattering is reflected in the analytic structure of celestial four-point amplitudes as a function of $\beta.$

To gain intuition about the structure of celestial four-point scattering, it is useful to consider some examples~\cite{Arkani-Hamed:2020gyp}. While the form of $\mathcal{M}(s, t)$ for all $s, t$ is unknown in general, we can extract some generic features by studying the structure of celestial amplitudes associated with $\mathcal{M}(s, t)$ polynomial and exponentially suppressed as  functions of $s$ at fixed scattering angle,  
\begin{equation} 
\mathcal{M}_{poly}(s, t) = s^a, \quad \mathcal{M}_{exp}(s, t) = e^{-\alpha s}.
\end{equation}
As anticipated, in the first case we find that the celestial amplitude is ill-defined for any $a \in \mathbb{R}$, while for  purely imaginary $\beta$ and $a$ it takes the form 
\begin{equation} 
\widetilde{\mathcal{A}}_{4,poly}(\Delta_i, z_i, \bar{z}_i) \propto \int_0^{\infty} d\omega \omega^{\beta - 1} \mathcal{M}_{poly}(\omega^2, - \omega^2 r) \propto \delta(\beta + 2a), \quad \beta, a \in i \mathbb{R}.
\end{equation}
In the second case, we find 
\begin{equation} 
\label{eq:exp-supr}
\widetilde{\mathcal{A}}_{4,exp}(\Delta_i, z_i, \bar{z}_i) \propto \int_0^{\infty} d\omega \omega^{\beta - 1} \mathcal{M}_{exp}(\omega^2, - \omega^2 r) \propto \alpha^{-\beta/2} \Gamma(\beta/2), \quad {\rm Re}(\alpha), {\rm Re}(\beta) > 0.
\end{equation}

Neither example can represent physical amplitudes as they violate basic consistency conditions~\cite{Adams:2006sv}, however both polynomial and exponential behaviours are expected to govern four-point scattering in certain theories and particular energy regimes, for instance in gravity at low and high energies respectively. It is then convenient to introduce an arbitrary soft-hard cutoff $\omega_*$ and split the Mellin integral in \eqref{eq:gca} as\footnote{~The cutoff dependence drops out in the full result upon adding up the high- and low-energy integrals. }
\begin{equation}
\label{eq:split-ca}
\widetilde{\mathcal{A}}_4(\Delta_i, z_i, \bar{z}_i) \propto \int_0^{\omega_*} d\omega \omega^{\beta - 1}\mathcal{M}(\omega^2, -\omega^2 r) +   \int_{\omega_*}^{\infty} d\omega \omega^{\beta - 1} \mathcal{M}(\omega^2, - \omega^2 r).
\end{equation}
For $\beta$ in the left hand complex plane, the second contribution in~\eqref{eq:split-ca} can be seen to be analytic provided that $\mathcal{M}(\omega^2, -\omega^2 r) \rightarrow 0$ as $\omega \rightarrow \infty$~\cite{Arkani-Hamed:2020blm}. As for the first contribution in~\eqref{eq:split-ca} note that at low energies and suppressing massless loops, $\mathcal{M}$ admits an expansion \cite{Arkani-Hamed:2020blm}
\begin{equation}
\mathcal{M}(s, t) = \sum_{m,n} a_{m, n} s^{n - m} t^m \iff \mathcal{M}(\omega^2, - \omega^2r) = \sum_{m, n} a_{m, n} \omega^{2n} (-r)^m, \quad \omega \leq \omega_*.
\end{equation}
Evaluating the first integral in~\eqref{eq:split-ca} using this expansion we thus see that celestial amplitudes have an infinite number of poles at negative even integer $\beta = -2n$.  Moreover, the residues of these poles are in one-to-one correspondence with coefficients in a low-energy expansion of momentum space scattering amplitudes or equivalently, couplings of higher dimension operators in a low-energy effective field theory description.

For $\beta$ in the right hand complex plane, the first contribution in~\eqref{eq:split-ca} is analytic while the second yields an infinite series of poles at positive even integer $\beta$ provided that $\mathcal{M}$ admits a series expansion around $\omega \rightarrow \infty.$ No such poles are present for exponentially suppressed amplitudes at high energies and fixed scattering angle as can be seen from \eqref{eq:exp-supr}.  We therefore conclude that exponentially soft behaviour at high energies leads to the absence of poles in massless celestial four-point scattering in the right-hand complex $\beta$ plane. Such behaviour is characteristic for theories of quantum gravity where black holes are expected to be responsible for the exponential suppression at high energies \cite{osti_5148846}. It is also a feature of string amplitudes where the tower of massive string states leads to similar behaviour already at tree level~\cite{GROSS1987129}. The high-energy properties of bulk scattering are reflected in the large-$\beta$ limit of celestial amplitudes where the Mellin integral \eqref{eq:gca} becomes dominated by high energies. Interestingly, an analysis of tree-level celestial four-point amplitudes in string theory suggests that in this limit the celestial sphere becomes the string worldsheet \cite{Stieberger:2018edy}.

We conclude this section with a review of explicitly known celestial amplitudes. Three-point celestial amplitudes involving massive particles were computed in \cite{Pasterski:2016qvg, Lam:2017ofc}. At tree-level, four-point amplitudes of scalars, gluons and gravitons were worked out in \cite{Pasterski:2017ylz, Stieberger:2018edy, Nandan:2019jas}, four-point string amplitudes were discussed in~\cite{Stieberger:2018edy}, while three- and four-point superamplitudes were computed in \cite{Brandhuber:2021nez, Jiang:2021xzy, Ferro:2021dub}. For certain classes of tree-level celestial amplitudes all-multiplicity formulas already exist in the literature and can be expressed in terms of generalised hypergeometric functions: Yang-Mills MHV and NMHV amplitudes were calculated in~\cite{Schreiber:2017jsr}, while~\cite{Casali:2020uvr} used the ambitwistor string to compute celestial amplitudes for biadjoint scalars, Yang-Mills and gravity. Finally, celestial loop amplitudes were found in \cite{Banerjee:2017jeg, Gonzalez:2020tpi}.

\subsection{Double Copy for Celestial Amplitudes}

Amplitudes in gravity and gauge theories obey remarkable relations known as double copy~\cite{Bern:2010ue} which state that gravitational amplitudes can be obtained by a well-defined ``squaring'' of gauge theory amplitudes.
These relations are known to hold to all multiplicities at tree-level~\cite{BjerrumBohr:2009rd,Stieberger:2009hq,Feng:2010my,Bern:2010yg}, are implied by string theory relations~\cite{Plahte:1970wy,Kawai:1985xq,BjerrumBohr:2010hn,BjerrumBohr:2010zs,BjerrumBohr:2009rd,Stieberger:2009hq,Tourkine:2016bak,Hohenegger:2017kqy,Casali:2019ihm,Casali:2020knc,Vanhove:2018elu,Vanhove:2020qtt} in their low energy limit, and also hold at loop-level in a plethora of cases and for many pairs of field theories (see~\cite{Bern:2019prr} and the SAGEX review Chapter~2 \cite{Bern:2022wqg} for comprehensive reviews). These relations heavily exploit translation invariance which is no longer manifest in the conformal basis. 
Nevertheless, features of amplitudes that are expected to reflect fundamental properties of the perturbative regime of quantum field theory should survive a change of basis. It is thus reasonable to expect a version of the double copy to exist for celestial amplitudes.

Consider the simplest case of tree-level scattering in Yang-Mills theory and Einstein gravity. In a basis of momentum eigenstates, the statement is that upon replacing the colour factors obeying the Lie algebra Jacobi identity $ c_s-c_t+c_u=0$ by colour-kinematics dual numerators obeying the same identities $ n_s-n_t+n_u=0$~\cite{Bern:2008qj}, the $n$-point Yang-Mills amplitude maps to an amplitude for $n$ external gravitons~\cite{Bern:2010yg} 
\begin{equation}\label{eq:doublecopy}
 \calA^{YM}_n=\delta^{(4)}\Big(\sum_{j=1}^n p^\mu_j\Big)\sum_{\gamma\in\Gamma_n} \frac{c_\gamma n_\gamma}{\Pi_\gamma}\quad \mapsto \quad \calA^{G}_n= \delta^{(4)}\Big(\sum_{j=1}^n p^\mu_j\Big)\sum_{\gamma\in\Gamma_n} \frac{n_\gamma^2}{\Pi_\gamma}\,.
\end{equation}
Here $\Gamma_{n}$ is the set of $n$-point trivalent graphs and the denominator $\Pi_\gamma$ is a product of the scalar propagators associated with the graph $\gamma$.
This presentation of the double copy relies on the fact that external particles are in the plane wave basis and on explicit momentum conservation. Both of these require a generalization in order to arrive at a double copy for celestial amplitudes. 

This can be achieved by considering a representation of the amplitudes obtained from position space Feynman diagrams where spacetime integrals like
\begin{equation}\label{eq:deltaInt}
 \delta^{(4)}\Big(\sum_{j=1}^n p_j^\mu\Big)=\int \frac{d^4 X}{(2\pi)^4} e^{\sum_{j=1}^n i p_j \cdot X}
 \end{equation} 
are left undone. Celestial amplitudes in Yang-Mills theory and gravity are then obtained from~\eqref{eq:doublecopy} by using~\eqref{eq:deltaInt} and replacing the plane waves $e^{i p_j\cdot X}$ by their Mellin transforms~\eqref{eq:phi} and the momenta $p^\mu_j=\pm \omega_j q^\mu_j$ by the translation generator of the Poincar\'{e}  algebra  
\begin{equation}\label{eq:Pmu}
 \calP^\mu_j = \pm q^\mu_j e^{\partial_{\Delta_j}}
\end{equation}
acting on the $j$-th conformal primary wavefunction. The colour-kinematics dual numerators $n_\gamma$ in~\eqref{eq:doublecopy} are thus promoted to {\it operator-valued} numerators $\calN_\gamma$.
This yields the {\it celestial double copy}~\cite{Casali:2020vuy}
\begin{equation}\label{eq:BCJ_cel}
\widetilde \calA^{YM}_n=\sum_{\gamma\in\Gamma}c_\gamma\mathcal{N}_{\gamma}S_\gamma \quad \mapsto  \quad \widetilde \calA^G_n=\sum_{\gamma\in\Gamma}(\mathcal{N}_{\gamma})^2S_\gamma \,,
\end{equation}
where the colour factors $c_\gamma$ are replaced by the operator-valued numerators $\calN_\gamma$, and $S_\gamma$ is the scalar amplitude for the trivalent graph $\gamma$.
An explicit discussion of~\eqref{eq:BCJ_cel} for four-point amplitudes is given in~\cite{Casali:2020vuy}. The all-multiplicity generalization is straightforward albeit cumbersome; an elegant proof using the ambitwistor string was given in~\cite{Casali:2020uvr}.
Note that the above presentation of celestial amplitudes is reminiscent of recent results on ambitwistor strings in Anti-de Sitter spacetime~\cite{Roehrig:2020kck,Eberhardt:2020ewh} which also has operator-valued kinematical numerators. It is suggestive that the natural way to implement the double copy in curved spacetimes will be through operator-valued numerators.

\subsection{OPEs from Collinear Limits of Celestial Amplitudes}
\label{sec:OPE-from-coll-lims}

Interesting and useful constraints on amplitudes can be inferred from their behaviour in special limits of the external momenta.
When two or more external momenta become collinear, tree-level massless momentum-space amplitudes factorise into a universal prefactor and lower point amplitudes as discussed in section~\ref{sec:sub-leading-st}. 
In terms of the ``celestial parametrization" of the momenta 
of section~\ref{subsec:holomap}, the collinear limit  $p_i||p_j$ corresponds to $z_i \rightarrow z_j$. In CCFT, when two operators associated to two gauge bosons are inserted near the same point on the celestial sphere, the singularities at $z_i=z_j$ correspond to the singularities of the operator product expansion (OPE). Therefore the collinear limit of momentum-space amplitudes extracts the celestial OPE. To match onto the literature we use the momenta~\eqref{eq:pi} rescaled by $\frac{1}{\sqrt{2}}$ as in~\cite{Pate:2019lpp} in all formulas involving OPEs.

Collinear divergences originate from the propagator poles $\frac{1}{(p_i+p_j)^2}\sim \frac{1}{p_i\cdot p_j}$ and have a universal form that can be derived from three-point vertices. 
In the following we will work in $(2,2)$ signature and consider the ``holomorphic" limit, $z_{ij}\equiv z_i -z_j\to 0$ with $\bz_i,\bz_j$ fixed, in which the momentum space amplitude becomes
\begin{equation}\label{eq:Acollinear}
 \lim_{z_{ij}\to 0} \calA_{\ell_1,\dots ,\ell_n}(p_1,\dots, p_n)=\sum_{\ell\in \pm s} {\rm Split}^\ell_{\ell_i ,\ell_j} (p_i,p_j)  \calA_{\ell_1,\dots ,\ell,\dots ,\ell_n}(p_1,\dots P,\dots, p_n)+\mathcal O(z_{ij}^{0}),
\end{equation}
where the combined momentum of the collinear pair is
\begin{equation}
 P^\mu=p^\mu_i + p^\mu_j=\omega_P q^\mu_P\,, \quad \omega_P=\omega_i+\omega_j
\end{equation}
and the splitting functions are given in \eqref{eq:split}.
To extract the OPE we compute the Mellin transform in the collinear limit~\eqref{eq:Acollinear}
\begin{equation}
 \prod_{k=1}^n  \int_0^\infty \frac{d\omega_k}{\omega_k} \omega_k^{\Delta_k}   \calA_{\ell_1,\dots ,\ell_n}(p_1,\dots, p_n)\stackrel{i||j}{\longrightarrow}
 \lim_{z_{ij}\to 0}\tcalA_{J_1,\dots ,J_n}(\Delta_1,z_1,\bz_1,\dots, \Delta_n,z_n,\bz_n) \,.
\end{equation}
Upon changing variables
\begin{equation}\label{eq:omegaP}
 \omega_i=\alpha\omega_P\,, \quad \omega_j=(1-\alpha)\omega_P,
\end{equation} 
the $\alpha$~integral is immediately recognizable as the integral representation of the Euler beta function
\begin{equation}\label{eq:EulerBeta}
 B(x,y)=\int_0^1 d\alpha \,\alpha^{x-1}(1-\alpha)^{y-1}=\frac{\Gamma(x)\Gamma(y)}{\Gamma(x+y)}\,,
\end{equation}
whose arguments are functions of $\Delta_i$ and $\Delta_j$.
Since the only $\alpha$-dependence comes from the splitting factor we can readily read off the celestial OPEs. 

We illustrate this for the OPEs of positive helicity outgoing gluons and gravitons.
%
%
The collinear splitting factors for positive helicity gluons takes the form~\cite{Bern:1998sv} 
\begin{equation}
 {\rm Split}^{{+}1}_{{+}1,{+}1}(p_i,p_j)= \frac{1}{z_{ij}} \frac{\omega_P}{\omega_i \omega_j}\,.
\end{equation}
Following the arguments above we can read off the celestial OPE~\cite{Fan:2019emx}
\begin{equation}\label{eq:gluonOPE}
\mathcal{O}^{a}_{\Delta_i,+1}(z_i, \bz_i) \mathcal{O}^{ b}_{\Delta_j,+1}(z_j, \bz_j) \sim -\frac{i f^{ab}_{~~c}}{z_{ij}} B(\Delta_i-1, \Delta_j-1) \mathcal{O}^{c}_{\Delta_i + \Delta_j - 1,+1}(z_j, \bz_j).
\end{equation}
%
%
For positive helicity gravitons the collinear splitting factor takes the form~\cite{Bern:1998sv} 
\begin{equation}
 {\rm Split}^{{+}2}_{{+}2,{+}2}(p_i,p_j)=-\frac{\kappa}{2} \frac{\bz_{ij}}{z_{ij}} \frac{\omega_P^2}{\omega_i \omega_j} 
\end{equation}
which gives the celestial OPE~\cite{Pate:2019lpp}
\begin{equation}
 \calO_{\Delta_i,+2}(z_i,\bz_i)\calO_{\Delta_j,+2}(z_j,\bz_j)\sim -\frac{\kappa}{2}\frac{\bz_{ij}}{z_{ij}} B(\Delta_i-1,\Delta_j-1)\calO_{\Delta_i+\Delta_j,+2}(z_j,\bz_j)\,.
\end{equation}
The celestial OPEs for mixed helicity gluons and gravitons, their incoming versions and the generalization of these results to account for the presence of both incoming and outgoing particles can be found in~\cite{Pate:2019lpp}. Celestial OPEs involving the two-dimensional stress tensor and other currents were discussed in~\cite{Fotopoulos:2019tpe,Fotopoulos:2019vac,Fotopoulos:2020bqj}, while OPEs involving particles of arbitrary spin were worked out in \cite{Himwich:2021dau}.

\section{Celestial Symmetries}
\label{sec:Celestial-symmetries}

In this section, we discuss the symmetries of celestial amplitudes and their implications. 

\subsection{Poincar\'e Symmetries}
Four-dimensional Lorentz symmetries act as global conformal symmetries on the celestial sphere \cite{Pasterski:2016qvg,Stieberger:2018onx}. By construction, the celestial operators defined in~\eqref{eq:2Dop} transform under Lorentz symmetries as global conformal primary operators in $2$D CFT, namely
\begin{equation} 
\label{eq:lg}
\begin{split}
[L_m, \mathcal{O}^{\pm}_{h, \bh}(z, \bz)] &= z^m\left((m + 1) h + z\partial_{z} \right)\mathcal{O}_{h, \bh}^{\pm}(z, \bz),\\
[\bar{L}_m, \mathcal{O}^{\pm}_{h, \bh}(z,\bz)] &= \bz^m \left((m + 1)\bh + \bz\partial_{\bz} \right) \mathcal{O}_{h, \bh}^{\pm}(z, \bz), 
\end{split}
\end{equation}
for $m = 0, \pm 1$.  Here $(h, \bh)$ are related to the conformal dimension $\Delta$ and spin $J$ by
\begin{equation} 
h = \frac{\Delta + J}{2}, \quad \bh = \frac{\Delta - J}{2}.
\end{equation}
These generators obey the SL$(2,\mathbb{C})$ algebra
\begin{equation}
\label{sl2c}
[L_m, L_n] = (m - n) L_{m + n}, \quad [\bar{L}_{m}, \bar{L}_{n}] = (m - n)L_{m+n}.
\end{equation}
Lorentz invariance of momentum-space scattering amplitudes implies that celestial \mbox{$n$-point} amplitudes obey the global conformal Ward identities
\begin{equation} 
\label{lc}
\sum_{j = 1}^n L_{m}^{(j)} \widetilde{\mathcal{A}}_n = \sum_{j = 1}^n \bar{L}_{m}^{(j)} \widetilde{\mathcal{A}}_n = 0.
\end{equation}

In contrast to Lorentz symmetries, translation symmetries act distinctly on in/out as well as massless and massive celestial operators. In the massless case, translations act simply as weight-shifting operators \cite{Stieberger:2018onx} 
\begin{equation}
\label{eq:ta}
[P_{k,l}, \mathcal{O}_{h, \bh}^{\pm}(z, \bz)] = \pm z^{k + \frac{1}{2}}\bz^{l + \frac{1}{2}} \mathcal{O}_{h + \frac{1}{2}, \bh + \frac{1}{2}}^{\pm},
\end{equation}
for $k,l = \pm \frac{1}{2}$. These are defined in terms of the components of~\eqref{eq:Pmu} as
\begin{equation}
 P_{\mp \frac{1}{2},\mp\frac{1}{2}}=\frac{1}{2}(\calP^0\pm \calP^3)\,, \quad P_{\pm \frac{1}{2},\mp \frac{1}{2}}=\frac{1}{2}(\calP^1\pm i\calP^2)\,.  
\end{equation}
The shift in the scaling dimension arises in a conformal primary basis from multiplication by a factor of energy in momentum space. Translation invariance therefore implies a relation among celestial amplitudes with shifted weights, 
\begin{equation} 
\label{eq:mc}
\sum_{j = 1}^n P_{-\frac{1}{2}, -\frac{1}{2}}^{(j)} \widetilde{\mathcal{A}}_n = 0.
\end{equation}
Under Lorentz symmetries, $P_{k,l}$ transform as 
\begin{equation}
\label{eq:LP}
    [L_m, P_{k,l}] = \left(\frac{m}{2} - k\right)P_{k + m, l}, \quad  [\bar{L}_m, P_{k,l}] = \left(\frac{m}{2} - l\right)P_{k, l+m}.
\end{equation}
Note that this is the same as the transformation property of the modes of a conformal primary operator of weights $(\frac{3}{2}, \frac{3}{2})$ \cite{Fotopoulos:2019vac}. 
Together with 
\begin{equation} 
\label{eq:PP}
[P_{k,l}, P_{k',l'}] = 0,
\end{equation}
equations~\eqref{sl2c} and~\eqref{eq:LP} define the Poincar\'e algebra. 

The action of translations on massive celestial operators of arbitrary spin can be found in~\cite{Law:2020tsg}, while the associated symmetry constraints have been discussed in~\cite{Law:2019glh,Law:2020tsg}.

\subsection{Conformally Soft Symmetries}
\label{subsec:css}

Gravity and gauge theories in four-dimensional asymptotically flat spacetimes enjoy a much larger symmetry group than Poincar\'e. 
For the case of gravity discussed in section~\ref{sec:st-as}, the leading and sub-leading soft graviton theorems imply that translations and Lorentz transformations are promoted to local (angle-dependent) symmetries -- supertranslations and superrotations respectively. Together, these form the $\mathfrak{bms}_4$ algebra defined by \eqref{sl2c}, \eqref{eq:LP} and \eqref{eq:PP} with $m, n \in \mathbb{Z}$ and $k,l \in \mathbb{Z} + \frac{1}{2}$ \cite{Barnich:2017ubf,Fotopoulos:2019vac}  .
The action of these symmetries on celestial operators is again given by~\eqref{eq:lg} and~\eqref{eq:ta}. As we will see, this algebra can also be extracted from OPEs of celestial amplitudes involving conformally soft gravitons \cite{Fotopoulos:2019vac, Banerjee:2020kaa, Banerjee:2020zlg, Banerjee:2020vnt, Guevara:2021abz, Banerjee:2021cly, Himwich:2021dau}. Supersymmetric generalizations of this algebra were studied in \cite{Fotopoulos:2020bqj}.

\subsubsection{Conformally Soft Operators}

The charges generating asymptotic symmetries at null infinity can be recast as special boundary operators in celestial CFT. 
For certain values of $\Delta \in \mathbb{Z}$ they correspond to soft charges~\cite{Donnay:2018neh,Donnay:2020guq} in the full (matter-coupled) theory when the Cauchy slice on which they are defined is taken to null infinity and the wavefunctions $\Phi_{\Delta,J}$ are the Goldstone modes of the spontaneously broken asymptotic symmetries. Indeed, the operators~\eqref{eq:2Dop} generate the shift
\begin{equation}\label{shift2}
[\calO^{\pm}_{\Delta,J}(z,\bz),\hat O^{s}(X)]=i\Phi^{s}_{\Delta,J}(X_\mp;z,\bz)
\end{equation}
for $s\in \mathbb{Z}_{\geq 0}$ expected for asymptotic symmetries in gauge theory and gravity. In supersymmetric theories for $s\in \frac{1}{2}\mathbb{Z}_{\geq0}$ we replace $[.\,,.]\mapsto \{.\,,.\}$ in~\eqref{shift2} and soft charges are found at special values of $\Delta \in\frac{1}{2} \mathbb{Z}$~\cite{Fotopoulos:2020bqj,Pano:2021ewd}. The celestial currents generating asymptotic symmetries discussed in the following are summarized in table~\ref{tab:Symmetries}.
\renewcommand{\arraystretch}{1.3}
\begin{table}
 \centering
\begin{tabular}{ |c|c|c|c|c|c| } 
 \hline
$s=|J|$ & $\Delta$ & $\tDelta$ & soft pole & celestial current & asymptotic symmetry \\ \hline
 $1$ & $1$ & $1$ &$\omega^{-1}$ & $J_z$  & large U(1) \\
 $\frac{3}{2}$ & $\frac{1}{2}$ & $\frac{3}{2}$ & $\omega^{-\frac{1}{2}}$ &  $\widetilde S_\bz$, $S_z$  & large SUSY\\ 
 $2$ & $1$ & $1$ &$\omega^{-1}$ & $P_z$  & supertranslation \\
$2$ & $0$ & $2$ &$\omega^{0}$ & $\widetilde T_{\bz\bz}$, $T_{zz}$  & superrotation \\
 \hline
\end{tabular}
\caption{Asymptotic symmetries generated by celestial currents. 
}
\label{tab:Symmetries}
\end{table}

\paragraph{Gauge Theory}

The 2D generators of large gauge symmetry are the operators~\eqref{eq:2Dop} with $\Delta = 1, J = \pm 1$ \cite{Donnay:2018neh, Pate:2019mfs}. These are obtained from 4D gauge fields according to~\eqref{eq:2Dop}, where the conformal primary wavefunction reduces to a pure gauge Goldstone wavefunction~\cite{Donnay:2018neh}
\begin{equation}\label{eq:A1}
    A_{\Delta = 1, J ;\mu}=  \nabla_{\mu} \Lambda_{gauge}\,, \quad \Lambda_{gauge}=-\frac{1}{\sqrt{2}}\partial_a \log\left( -q \cdot X \right),
\end{equation}
where $a = z ~(\bar{z})$ for $J = +1 ~(-1)$. 
The asymptotic value of the gauge parameter $\Lambda_{gauge}$ at null infinity is proportional to the special choice of the function $\varepsilon$ from section~\ref{sec:st-as} which establishes the equivalence between the Ward identity of large gauge symmetry and the soft photon theorem. The conserved operator is the {\it conformally soft} photon current $J_z$ with $(h,\bh)=(1,0)$, or $J_\bz$ with $(h,\bh)=(0,1)$.
\paragraph{Gravity}
The construction of the BMS supertranslation current $P_z$ ($P_\bz$) is similar and related to a celestial operator of $\Delta = 1, J =  +2 ~(J=-2)$ associated with the Goldstone wavefunction~\cite{Donnay:2018neh}
\begin{equation}\label{eq:h1}
  h_{\Delta = 1, J;\mu\nu} = \nabla_{\mu} \xi_{\nu, a} + \nabla_{\nu} \xi_{\mu, a}=\nabla_\mu \nabla_\nu \Lambda_{gravity},\quad \Lambda_{gravity}=\frac{1}{4} \partial_a^2[(-q\cdot X)\log(-q\cdot X)], 
\end{equation}
with $a = z ~(\bar{z})$ for $J = +2 ~(-2)$. 
The asymptotic value of $\Lambda_{gravity}$ is related to the supertranslation parameter $f$ appearing in the BMS charge conservation law in section~\ref{sec:st-as} associated to the leading soft graviton theorem. The Diff($S^2$) superrotation generators are associated with $\Delta = 0,~ J = \pm 2$ pure gauge conformal primary wavefunctions which can also be expressed as diffeomorphisms~\cite{Donnay:2020guq}.  
The operator obtained from the $\Delta=0$ mode (denoted $\widetilde T_{\bz\bz}$ in table~\ref{tab:Symmetries}) is related  by a shadow transform~\cite{Donnay:2020guq} to the  $\widetilde\Delta=2-\Delta=2$ stress tensor $T_{zz}$ defined in~\eqref{eq:Tzz-ward} with $(h,\bh)=(2,0)$ that generates a Virasoro symmetry. The $\widetilde \Delta=2$ diffeomorphism is expressible in terms of the  meromorphic vector field $Y^A$ used in establishing the relation between the sub-leading soft graviton theorem and celestial conformal symmetry.
\bigskip

The asymptotic symmetries of gauge theory and (super)gravity\footnote{~A similar analysis for large supersymmetry transformations~\cite{Lysov:2015jrs,Avery:2015iix} related to the soft gravitino theorem can be found in~\cite{Pasterski:2020pdk,Pano:2021ewd} which is generated by the $\widetilde\Delta=\frac{3}{2}$ supercurrent $S_z$ or its $\Delta=\frac{1}{2}$ shadow $\widetilde S_{\bz}$.} can thus be viewed as being generated by two-dimensional currents in CCFT. From the Kerr-Schild double copy relation between the conformal primaries~\eqref{eq:CPWs} we see for $\Delta=1$ that BMS symmetry can be regarded as a double copy of large gauge symmetry~\cite{Huang:2019cja,Alawadhi:2019urr}. 
For $\Delta \in 1-\mathbb{Z}_{\geq0}$ there exists an infinite tower of conformally soft primaries obeying a Kerr-Schild double copy~\cite{Pasterski:2020pdk}; see also~\cite{Campiglia:2021srh}.

\subsubsection{Conformally Soft Theorems and 2D Ward Identities}

Another way to see the action of asymptotic symmetries in celestial CFT is to derive the imprint of soft theorems on celestial amplitudes. It is not immediately obvious how to reveal the expected universal behaviour of celestial amplitudes in the soft limit, or equivalently, how to take a ``low energy'' limit of a boost eigenstate which involves a superposition of all energy eigenstates. Nevertheless, motivated by the analysis of conformal primary wavefunctions of $\Delta = 0, 1$ and the discussion in section \ref{sec:analytic-structure} 
we can start with a massless celestial operator of dimension $\Delta$ and spin $J$ given by
\begin{equation} 
\mathcal{O}_{\Delta, J}(z, \bz) = \int_0^{\infty}d\omega \omega^{\Delta - 1} \mathcal{O}_{\ell= J}(\omega, z, \bz),
\end{equation}
and consider the limit 
\begin{equation} 
\begin{split}
\lim_{\Delta \rightarrow -n} (\Delta + n) \mathcal{O}_{\Delta,J}(z, \bz) 
&= \lim_{\Delta \rightarrow -n}(\Delta + n) \sum_{k} \int_0^{\omega_*}d\omega \omega^{\Delta + k - 1}  O_{J,k}(z, \bz) =  O_{J,n}(z, \bz).
\end{split}
\end{equation}
Here we expanded
\begin{equation} 
\mathcal{O}_J(\omega, z, \bz) = \sum_k \omega^k  O_{J,k}(z, \bz)
\end{equation}
 for $\omega \leq \omega_*$ and assumed that insertions of $\mathcal{O}_J(\omega, z, \bz)$ into $\mathcal{S}$-matrices have fast enough fall-offs with energy\footnote{~An exponential fall-off $\lim_{\omega \rightarrow \infty}\langle \mathcal{O}(\omega, z, \bz)\cdots \rangle \sim e^{-\epsilon \omega}$ will ensure this limit is well defined for any negative integer $\Delta$.} in which case the  high-energy part of the Mellin integral will be free of poles in $\Delta + n$. We therefore see that the $\Delta \rightarrow -n$ limit of a celestial operator for $n = -1, 0, 1,...$ picks out the $\mathcal{O}(\omega^n)$ term in an expansion around $\omega = 0$. The $\Delta = 1, 0, -1$ insertions are associated with leading, sub-leading and sub-sub-leading soft behaviour respectively~\cite{Fan:2019emx,Nandan:2019jas,Pate:2019mfs,Adamo:2019ipt,Puhm:2019zbl,Guevara:2019ypd}. 

In the following sections we review the universal behaviour of gravity and gauge theory celestial amplitudes in conformally soft limits, as well as in the collinear or OPE limit. The analysis will uncover interesting infinite dimensional symmetry algebras \cite{Guevara:2021abz, Strominger:2021mtt} which will be the subject of section~\ref{ssec:winf}.

\paragraph{Gauge Theory}
It was shown in \cite{He:2014cra} that, in theories without massive particles, the leading soft photon theorem implies that soft photons behave as U(1) currents. 
To see this, one parameterizes the massless momenta as in \eqref{eq:masslessmom} and substitutes them into the leading soft photon relation \eqref{eq:soft-photon-factors}. For outgoing soft photons of momentum $p$ and hard particles of momenta $p_k$ (again rescaled as in section \ref{sec:OPE-from-coll-lims}) using 
\begin{equation} 
\begin{split}
p_k \cdot \epsilon_{+}(z, \bar{z}) 
= - \omega_k \eta_k (\bar{z} - \bar{z}_k), \quad
p_k \cdot p(z, \bar{z}) = - \omega  \omega_k \eta_k |z - z_k|^2,
\end{split}
\end{equation}
 the soft photon factor becomes
\begin{equation} 
S^{(0)}_+ = \sum_{k = 1}^n \frac{Q_k}{\omega(z - z_k)},
\end{equation}
where we absorbed the factor of $\eta_k$ into the definition of $Q_k$. 
Consequently, $\mathcal{S}$-matrices with soft photon insertions obey Ward identities of the form 
\begin{equation} 
\label{u1c}
\begin{split}
\langle J_z(z, \bz) \prod_{i=1}^n \mathcal{O}_i(\omega_i, z_i, \bz_i)\rangle &\equiv \lim_{\omega \rightarrow 0}\omega \langle \mathcal{O}_{\ell = 1}(\omega, z, \bz) \prod_{i=1}^n \mathcal{O}_i(\omega_i, z_i, \bz_i)\rangle\\
&= \sum_{k = 1}^n \frac{Q_k}{z - z_k} \langle \prod_{i=1}^n \mathcal{O}_i(\omega_i, z_i, \bz_i) \rangle.
\end{split}
\end{equation}
Similarly, the leading soft gluon theorem can be recast as a holomorphic Kac-Moody symmetry generated by non-abelian currents $J_z^a$ (or soft gluons of positive helicity)~\cite{He:2015zea}.

As explained in section~\ref{subsec:css}, in a conformal primary basis soft photons or gluons correspond to operators of vanishing holomorphic/antiholomorphic weights or equivalently $\Delta = 1$ and $J = \pm 1$. The conformally soft theorem then relates celestial amplitudes with and without insertions of such operators, namely 
\begin{equation} 
\label{eq:csth-ph}
\lim_{\Delta \rightarrow 1} (\Delta - 1)\widetilde{\mathcal{A}}_{n+1}(\Delta, J = +1, z,\bz; \Delta_i, z_i, \bz_i) = \sum_{k = 1}^n \frac{Q_k}{z - z_k} \widetilde{\mathcal{A}}_n(\Delta_i, z_i, \bz_i).
\end{equation}
This is the celestial counterpart of the soft photon theorem~\eqref{eq:sphth} as first derived in~\cite{Fan:2019emx} and~\cite{Nandan:2019jas,Pate:2019mfs}. Its non-abelian gauge theory analog was verified in examples of celestial amplitudes in Yang-Mills and open string theory in~\cite{Pate:2019mfs}. 

In the same parameterizations for the momenta, the sub-leading soft photon factor \eqref{eq:soft-photon-factors} takes the form 
\begin{equation}
    S^{(1)}_+ = \sum_{k = 1}^n \frac{Q_k \eta_k}{\omega_k(z-z_k)} \left(\ell_k + \omega_k\partial_{\omega_k} + (\bz - \bz_k)\partial_{\bz_k} \right),
\end{equation}
or in a conformal primary basis 
\begin{equation} 
 \widetilde{S}^{(1)}_+ = \sum_{k = 1}^n \frac{Q_k}{(z-z_k)} \left(-2\bar{h}_k + 1 + (\bz - \bz_k)\partial_{\bz_k} \right)(P_{-\frac{1}{2},-\frac{1}{2}}^{(k)})^{-1}.
\end{equation}
Here $P^{-1}_{-\frac{1}{2}, -\frac{1}{2}}$ is the inverse of the momentum mode \eqref{eq:ta}, 
\begin{equation}
   [P^{-1}_{-\frac{1}{2}, -\frac{1}{2}},\mathcal{O}^{\pm}_{h, \bar{h}}(z, \bz)] = \pm \mathcal{O}^{\pm}_{h - \frac{1}{2}, \bar{h} - \frac{1}{2}}(z, \bz). 
\end{equation}
The sub-leading conformally soft theorem then implies a differential recursion relation for celestial amplitudes
\begin{equation} 
\label{eq:sub-leading-recursion}
\begin{split}
\lim_{\Delta \rightarrow 0} &\Delta \widetilde{\mathcal{A}}_{n + 1}(\Delta, J = +1, z,\bz; \Delta_i, z_i, \bz_i) \\
&= \sum_{k = 1}^n \frac{Q_k}{(z-z_k)} \left(-2\bar{h}_k + 1 + (\bz - \bz_k)\partial_{\bz_k} \right)(P_{-\frac{1}{2},-\frac{1}{2}}^{(k)})^{-1} \widetilde{\mathcal{A}}_n(\Delta_i, z_i, \bz_i).
\end{split}
\end{equation}
We will see in section~\ref{sec:OPE-from-symm} how \eqref{eq:sub-leading-recursion} and their analog in non-abelian gauge theories can be used to constrain the dynamics encoded in OPE coefficients of celestial operators.  

\paragraph{Gravity}
The conformally soft behaviour of gravity amplitudes follows similarly by computing residues of poles at integer $\Delta \leq 1.$ In particular, in a conformal primary basis the leading soft graviton theorem \eqref{eq:lsgt} becomes \cite{Adamo:2019ipt,Puhm:2019zbl}
\begin{equation}
    \label{eq:csth-gr}
\lim_{\Delta \rightarrow 1} (\Delta - 1)\widetilde{\mathcal{A}}_{n+1}(\Delta, J = +2, z,\bz; \Delta_i, z_i, \bz_i) = -\frac{\kappa}{2}\sum_{k = 1}^n \frac{\bz - \bz_k}{z - z_k} \eta_k \widetilde{\mathcal{A}}_n(\Delta_k + 1, z_k, \bz_k),
\end{equation}
where all suppressed arguments on the right-hand side are unshifted. The shift in the $k$-th conformal dimension is inherited from multiplication by $\omega_k$ in a momentum-space basis. 

At sub-leading order, the conformally soft theorem is related to \eqref{eq:Tzz-ward} by a shadow transform \cite{Kapec:2016jld,Pasterski:2017kqt,Donnay:2018neh} and simply reflects the conformal symmetry of celestial amplitudes. An expression for the sub-sub-leading soft graviton theorem can be found in \cite{Campiglia:2016jdj,Campiglia:2016efb,Pate:2019lpp} and can be used to constrain the leading OPE behaviour of gravity amplitudes \cite{Pate:2019lpp} as we will see next. The celestial counterpart of the soft graviton theorem to sub-leading and sub-sub-leading order was discussed in~\cite{Guevara:2019ypd}. At tree level, celestial amplitudes obey further Ward identities associated with poles at increasingly negative $\Delta$. These were recently worked out from a celestial CFT perspective in \cite{Guevara:2021abz, Himwich:2021dau, Jiang:2021xzy}, while their asymptotic symmetry interpretation was explained in \cite{Freidel:2021ytz, Freidel:2021dfs}. Corrections to these Ward identities from higher-derivative operators and their associated constraints were computed in \cite{Jiang:2021ovh,Mago:2021wje}.

\subsection{OPEs from Symmetry}
\label{sec:OPE-from-symm}

We now derive the celestial operator product expansions, obtained in the previous section from collinear limits of celestial amplitudes, purely from symmetry considerations.
We treat $z_{i}, \bz_{i}$ as real independent variables
and study OPE expansions of celestial operators in a holomorphic collinear limit $z_{12} \rightarrow 0$ for $\bar{z}_1, \bar{z}_2$ fixed. We focus on operators creating outgoing particles and omit the corresponding label.

\paragraph{Gauge Theory}
We start  by assuming that positive-helicity gluons admit the holomorphic collinear expansion
\begin{equation} 
\label{gOPE}
\mathcal{O}^{a}_{\Delta_1, +1}(z_1, \bz_1) \mathcal{O}^{b}_{\Delta_2, +1}(z_2, \bz_2) \sim -\frac{i f^{ab}_{~~c}}{z_{12}} C(\Delta_1, \Delta_2) \mathcal{O}^{c}_{\Delta_1 + \Delta_2 - 1, +1}(z_2, \bz_2) + \dots,
\end{equation}
where $\dots$ include contributions from SL$(2, \mathbb{C})$ descendants. The form of the OPE is fixed by the leading soft theorem and SL$(2,\mathbb{C})$ up to a coefficient $C(\Delta_1, \Delta_2)$. We now show that the sub-leading conformally soft gluon theorem determines this leading OPE coefficient up to a normalisation fixed by the leading soft gluon theorem~\cite{Pate:2019lpp}. 

The negative helicity sub-leading soft gluon theorem leads to the following transformation properties 
\begin{equation} 
\begin{split}
\bar{\delta}_b \mathcal{O}^{a}_{\Delta, \pm 1}(z, \bz) = -(\Delta - 1 \mp 1 + \bz\partial_{\bz}) i f^a_{\ bc} \mathcal{O}^{c}_{\Delta - 1, \pm 1}(z,\bz)\,.
\end{split}
\end{equation}
Acting with $\bar{\delta}$ on both sides of \eqref{gOPE} and comparing the two sides, we deduce that $C(\Delta_1, \Delta_2)$ obey the recursion relation
\begin{equation} 
\label{rr}
(\Delta_1 - 2) C(\Delta_1 - 1, \Delta_2) = (\Delta_1 + \Delta_2 - 3) C(\Delta_1, \Delta_2)\,.
\end{equation}
\eqref{rr} has the unique\footnote{~By Wieland's theorem, see appendix E of \cite{Pate:2019lpp}. The normalisation is fixed by the leading soft theorem.} solution
\begin{equation} 
\label{eq:gluon-lOPE}
C(\Delta_1, \Delta_2) = B(\Delta_1 - 1, \Delta_2 - 1)\,.
\end{equation}
The OPE of opposite helicity gluons can be derived along similar lines and is given~\cite{Pate:2019lpp}.

\paragraph{Gravity}
A similar argument can be used to derive the leading behaviour of gravitons in the collinear limit. The OPE of positive helicity gravitons is fixed by SL$(2,\mathbb{C})$ to take the form
\begin{equation} 
\label{grav-OPE}
 \partial_{\bz_1}\mathcal{O}_{\Delta_1, +2}(z_1, \bz_1) \mathcal{O}_{\Delta_2, +2}(z_2, \bz_2) \sim \frac{D(\Delta_1, \Delta_2)}{z_{12}}  \mathcal{O}_{\Delta_1 + \Delta_2, +2}(z_2,\bz_2) + \dots,
\end{equation}
where $\dots$ include contributions from SL$(2,\mathbb{C})$ descendants, as well as primaries appearing at sub-leading order in a holomorphic collinear expansion. Here $\mathcal{O}_{\Delta_1 + \Delta_2, +2}$ is an SL$(2,\mathbb{C})$ primary, but a Poincar\'e descendant 
\begin{equation} 
\mathcal{O}_{\Delta_1 + \Delta_2, +2}(z) = P_{-\frac{1}{2}, -\frac{1}{2}} \mathcal{O}_{\Delta_1 + \Delta_2 - 1, +2}(z).
\end{equation}
Imposing that \eqref{grav-OPE} is invariant under the sub-sub-leading conformally soft action\footnote{~The associated asymptotic charges have been discussed in \cite{Freidel:2021dfs}.}
leads to a recursion relation 
for $D(\Delta_1, \Delta_2)$ that is solved by \cite{Pate:2019lpp}
\begin{equation} 
\label{eq:grav-lOPE}
D(\Delta_1, \Delta_2) = -\frac{\kappa}{2} B(\Delta_1 - 1, \Delta_2 - 1).
\end{equation}
Again, the normalisation is fixed by the leading conformally soft behaviour
\begin{equation}
\lim_{\Delta_1 \rightarrow 1} (\Delta_1  - 1) \mathcal{O}_{\Delta_1, +2}(z_1, \bz_1) \mathcal{O}_{\Delta_2, +2}(z_2, \bz_2) \sim -\frac{\kappa}{2}\frac{\bz_{12}}{z_{12}} \mathcal{O}_{\Delta_1 + \Delta_2, +2}(z_2,\bz_2).
\end{equation}

Both \eqref{eq:gluon-lOPE} and \eqref{eq:grav-lOPE} can be shown to holographically reproduce the collinear splitting functions derived in section \ref{sec:OPE-from-coll-lims}. More recently, it was shown that the same results follow upon including conformal descendant contributions to the OPEs and imposing Poincar\'e symmetry~\cite{Himwich:2021dau}.  

The graviton OPE expansion includes SL$(2,\mathbb{C})$ primary contributions at sub-leading orders in a $z_{12}$ expansion. Some of these were determined using the extended BMS symmetry algebra in \cite{Banerjee:2020zlg,Banerjee:2021cly} (see also \cite{Ebert:2020nqf} for an analysis of subleading terms in the gluon OPE). For example, at $\mathcal{O}(z_{12}^0\bz_{12}^0)$ SL$(2,\mathbb{C})$ implies that the positive helicity graviton OPE may receive contributions from operators of dimension $\Delta = \Delta_1 + \Delta_2$ and spin $J = 4.$ In pure gravity, the only SL$(2,\mathbb{C})$ primaries with these
dimensions are the two extended algebra descendants
\begin{equation} 
P_{-\frac{3}{2}, \frac{1}{2}}\mathcal{O}_{\Delta_1+ \Delta_2 - 1, +2}, \quad J^{1}_{-1}P_{-\frac{1}{2},-\frac{1}{2}}\mathcal{O}_{\Delta_1 + \Delta_2 -1, +2},
\end{equation}
where $J_{-1}^{1}$ is a mode of one of the current algebra generators extracted from the sub-leading conformally soft graviton which will be defined in \eqref{eq:conf-soft-grav-modes}. Specifically $J^1(z) 
\equiv H_{1}^0(z)$ and $J^1_{-1} = \oint dz z^{-1} J^1(z)$.
These operators can be shown to be related by a null state condition \cite{Banerjee:2020zlg}, therefore the graviton OPE gets corrected by only one term at sub-leading order,
\begin{equation} 
\label{gr-OPE-sub-leading}
\begin{split}
\mathcal{O}_{\Delta_1, +2}(z_1,\bz_1) \mathcal{O}_{\Delta_2, +2}(z_2, \bz_2) &\sim \frac{\bz_{12}}{z_{12}} D(\Delta_1,\Delta_2) P_{-\frac{1}{2}, -\frac{1}{2}}\mathcal{O}_{\Delta_1+ \Delta_2 - 1, +2}(z_2,\bz_2) \\
&+ D'(\Delta_1, \Delta_2) P_{-\frac{3}{2}, \frac{1}{2}}\mathcal{O}_{\Delta_1+ \Delta_2 - 1, +2}(z_2,\bz_2) + \dots.
\end{split}
\end{equation}
Since the sub-leading term is effectively a current algebra descendant of the first, its OPE coefficient should be determined by symmetry. Indeed, imposing invariance of \eqref{gr-OPE-sub-leading} under $J_1^{-1} = \oint dz z H^0_{-1}(z)$ and matching the $\mathcal{O}(z_{12}^0\bz_{12}^0)$ terms on both sides, one finds \cite{Banerjee:2020zlg} 
\begin{equation} 
D'(\Delta_1, \Delta_2) = -D(\Delta_1, \Delta_2). 
\end{equation}
MHV amplitudes can be shown to obey this relation \cite{Banerjee:2020zlg}. 

Similar arguments can be in principle used to determine further sub-leading terms in the OPE expansion of gluons and gravitons and can be checked to reproduce increasingly sub-leading behaviour of the corresponding MHV amplitudes \cite{ Banerjee:2020kaa,Banerjee:2020zlg,Ebert:2020nqf,Banerjee:2020vnt,Banerjee:2021cly,Banerjee:2021dlm}. Subtleties related to the mixing of helicity sectors appear in examples beyond MHV and generalizing these techniques to these cases is an active area of research. Recent progress in related fields such as twistor theory \cite{Adamo:2021lrv,Adamo:2021zpw,Costello:2022wso}  already appears to provide some insight into these problems, see also the SAGEX review Chapter~6 \cite{Geyer:2022cey}.

\subsection{Conformally Soft Symmetry Algebras} 
\label{ssec:winf}
We hinted in section~\ref{subsec:css} at an infinity of conformally soft operators 
\begin{equation}
    \lim_{\Delta \to -n} (\Delta+n) \calO_{\Delta,J}(z,\bz)\,, \quad 
 n=-1,0,1, \cdots.
\end{equation}
This limit effectively picks out the coefficients of the $\omega^n$ term in the expansion around $\omega \to 0$ of $\calO(\omega,z,\bz)=\sum_j \omega^j O_j(z,\bz)$.  As we will see in section~\ref{ssec:diamond}, the existence of these operators can also be derived from conformal representation theory. In this section we review the algebra of the \mbox{(semi-)}infinite tower of conformally soft gluon and graviton operators~\cite{Guevara:2021abz,Strominger:2021lvk,Himwich:2021dau}. We focus on only positive helicity currents
and work in a Vir${}_L\otimes$ SL(2,$\mathbb{R}$)${}_R$-invariant formalism where SL(2,$\mathbb{R}$)${}_R$ is the global subgroup of Vir${}_R$ superrotations. That is we treat $z$ and $\bz$ again as independent variables which amounts to continuing $(3,1)$ Minkowski to $(2,2)$ Klein space with the celestial sphere becoming the celestial torus \cite{Atanasov:2021oyu} and the Lorentz SL(2,$\mathbb{C})$ continued to SL(2,$\mathbb{R}$)${}_L\times$ SL(2,$\mathbb{R}$)${}_R$. 

\paragraph{Gauge Theory} 

Define the discrete family of positive helicity soft gluons
\begin{equation}\label{eq:Rsoft}
    R^{k,a}:=\lim_{\varepsilon \to 0} \varepsilon \calO^a_{k+\varepsilon,+1}\,, \quad k=1,0,-1,-2,...
\end{equation}
with weights
\begin{equation}
    (h,\bh)=\left(\frac{k+1}{2},\frac{k-1}{2}\right)\,,
\end{equation}
and a consistently-truncated antiholomorphic mode expansion
\begin{equation}
    R^{k,a}(z,\bz)=\sum_{n=\frac{k-1}{2}}^{\frac{1-k}{2}} \frac{R^{k,a}_n(z)}{\bz^{n+\frac{k-1}{2}}}\,.
\end{equation}
These values of conformal weights $\Delta=k$ include all the conformally soft poles encountered in the OPE~\eqref{eq:gluonOPE} of two positive helicity gluons. The factor of $\varepsilon$ is incorporated in~\eqref{eq:Rsoft} to cancel these poles and give finite OPEs for the rescaled $R^{k,a}$. Including the contribution from SL(2,$\mathbb{R}$)${}_R$ descendants the OPE~\eqref{eq:gluonOPE} becomes
\begin{eqnarray} 
\label{Oped}
\kern-60pt\mathcal{O}^{a}_{\Delta_1,+1}(z_1, \bz_1) \mathcal{O}_{\Delta_2,+1}^{ b}(z_2, \bz_2)& \sim \frac{-i f^{ab}_{~~ c}}{z_{12}} \sum_{n = 0}^{\infty} B(\Delta_1 - 1 + n, \Delta_2 - 1) \frac{\bz_{12}^n}{n!}\nonumber\\ &\qquad\qquad~~~~~~~~\times \bar{\partial}^n\mathcal{O}^{c}_{\Delta_1 + \Delta_2 - 1,+1}(z_2, \bz_2) + \dots
\end{eqnarray}
where $\dots$ denote terms sub-leading in the limit $z_{12} \rightarrow 0$. The pole structure of the Euler beta functions in \eqref{Oped} implies that for $\frac{k  -1}{2} \leq n \leq \frac{1 - k}{2}$, the mode operators $R^{k,a}_n(z)$ organize into $(2 - k)$-dimensional SL(2, $\mathbb{R}$)${}_R$ representations. In particular, note that $\bar{\partial}^{2 - k} R^{k, a}(z, \bz) = 0$. One can now infer the OPE of the soft currents~\eqref{eq:Rsoft} and derive the algebra of the soft operators~\cite{Guevara:2021abz}
\begin{equation} 
\label{eq:Ralgebra}
[R^{k, a}_n, R^{l, b}_{n'}] = -i f^{ab}_{~~ c}\left(\begin{matrix}
\frac{1 - k}{2} - n + \frac{1 - l}{2} - n'\\
\frac{1 - k}{2} - n
\end{matrix} \right) \left(\begin{matrix} \frac{1 - k}{2} + n + \frac{1 - l}{2} + n'\\
\frac{1 - k}{2} + n\end{matrix}\right) R^{k + l - 1, c}_{n + n'}.
\end{equation}
Defining the rescaled operators
\begin{equation}
\hat{R}^{p= \frac{3-k}{2}, a}_n \equiv {\textstyle\left(\frac{1 - k}{2} - n\right)! \left(\frac{1 - k}{2} + n\right)!} R^{k, a}_n,
\end{equation}
with $p = 1, \frac{3}{2}, 2, \dots $
~\eqref{eq:Ralgebra} implies $\hat{R}$ obey the simpler algebra
\begin{equation} 
\label{eq:Rhatalgebra}
[\hat{R}^{p, a}_n, \hat{R}^{q, b}_{n'}] = -i f^{ab}_{~~ c} \hat{R}^{p+q - 1, c}_{n + n'}.
\end{equation}

To summarize, in 4D non-abelian gauge theory with group $G$, the leading soft gluon theorem implies a standard closed celestial $G$-current algebra~\cite{Strominger:2013lka,Fan:2019emx,Pate:2019mfs,Adamo:2019ipt,Nandan:2019jas}. The sub-leading soft theorem implies two further $G$-valued holomorphic currents~\cite{Lysov:2014csa,Pate:2019lpp,Banerjee:2020vnt}  which form an SL(2,$\mathbb{R}$)${}_R$ doublet. In the conformal basis the leading and sub-leading soft gluons correspond, respectively, to the $\Delta=1$ and $\Delta=0$ conformally soft gluon operators. The commutator of two of these currents gives rise to further symmetry generators in an SL(2,$\mathbb{R}$)${}_R$ triplet. Continuing in this manner, there exists an infinite tower of $G$-currents in finite-dimensional SL(2,$\mathbb{R}$)${}_R$ representations obeying the closed algebra~\eqref{eq:Rhatalgebra}.

\paragraph{Gravity} 
A similar symmetry algebra analysis can be done for gravitons. 
Define a discrete family of conformally soft positive-helicity gravitons
\begin{equation}
 H^k:=\lim_{\varepsilon \to 0} \varepsilon \calO_{k+\epsilon,+2}\,, \quad k=2,1,0,-1,-2,...
\end{equation}
with weights
\begin{equation}
 (h,\bh)=\left(\frac{k+2}{2},\frac{k-2}{2}\right)\,,
\end{equation}
and a consistently-truncated antiholomorphic mode expansion
\begin{equation}
\label{eq:conf-soft-grav-modes}
 H^k(z,\bz)=\sum_{n=\frac{k-2}{2}}^{\frac{2-k}{2}}\frac{H_n^k(z)}{\bz^{n+\frac{k-2}{2}}}\,.
\end{equation}
As in gauge theory, the soft graviton currents $H^k$ obey a closed algebra~\cite{Guevara:2021abz,Strominger:2021lvk,Himwich:2021dau}. Following similar steps as above and redefining
\begin{equation}\label{eq:w}
 w_n^p=\frac{1}{\kappa}(p-n-1)!(p+n-1)! H_n^{-2p+4},
\end{equation}
one can show that the rescaled conformally soft graviton operators satisfy~\cite{Strominger:2021lvk}
\begin{equation}\label{eq:winf}
 [w_m^p,w_n^q]=[m(q-1)-n(p-1)]w_{m+n}^{p+q-2}\,,
\end{equation}
with $p,q$ running over positive half-integers $p,q=1,\frac{3}{2},2,\frac{5}{2},...$. This algebra is called the wedge algebra of $w_{1+\infty}$~\cite{Pope:1991ig} - `wedge' because of the restriction $ \frac{k-2}{2}\leq m \leq \frac{2-k}{2}$, or equivalently $1-p\leq m \leq p-1$. Evidence that such algebraic structures are encountered in a large $r$ expansion of asymptotically flat metrics obeying the vacuum Einstein equations was recently found \cite{Freidel:2021dfs,Freidel:2021ytz}. This algebra was shown to be uncorrected in self-dual gravity \cite{Ball:2021tmb}, while modifications of this algebra in the presence of non-minimal couplings and their implications for low-energy effective field theories were worked out in \cite{Mago:2021wje}. It will be interesting to explore the possible quantum deformations of this algebra and the constraints they impose on quantum theories of gravity.

\subsection{Global Conformal Multiplets in Celestial CFT}
\label{ssec:diamond}

We end the discussion of celestial symmetries with an examination of the structure of global conformal multiplets in two-dimensional celestial CFT. This will reveal the power of symmetry in organizing the conformally soft behaviour of scattering and unify the discussion of conformally soft theorems, two-dimensional Ward identities for four-dimensional asymptotic symmetries, their associated soft charges, and conformal dressings for celestial amplitudes.
Global conformal multiplets in celestial CFT were studied in~\cite{Pasterski:2021fjn} following the spirit of~\cite{Penedones:2015aga} from the bootstrap literature. Inspiration is drawn from the closely related discussion of null states in~\cite{Banerjee:2018gce,Banerjee:2018fgd,Banerjee:2019aoy,Banerjee:2019tam,Banerjee:2020kaa,Banerjee:2020zlg,Banerjee:2020vnt,Guevara:2021abz}. 

\begin{figure}[h]
\begin{center}
\vspace{-0.5em}
\begin{tikzpicture}[scale=1.3]
\definecolor{darkgreen}{rgb}{.0, 0.5, .1};
\draw[thick](0,2)node[above]{$(\frac{1-k}{2}, \frac{1-\bar k}{2}) $} ;
\draw[thick,->](0,2)--node[above left]{$\bar \partial^{ \bar k}$} (-1+.05,1+.05);
\draw[thick,->](0,2)--node[above right]{$\partial^{  k}$} (2-.05,0+.05);
\draw[thick,->] (-1+.05,1-.05)node[left]{$ (\frac{1-k}{2}, \frac{1+\bar k}{2}) $} --node[below left]{$\partial^{k}$} (1-.05,-1+.05) ;
\draw[thick,->] (2-.05,0-.05)node[right]{$(\frac{1+k}{2}, \frac{1-\bar k}{2}) $} --node[below right]{${\bar \partial}^{ \bar k}$} (1.05,-1+.05) ;
\draw[thick](1,-1.2)node[below]{$( \frac{1+k}{2}, \frac{1+\bar k}{2}) $};
\filldraw[black] (0,2) circle (2pt) ;
\filldraw[black] (2,0) circle (2pt) ;
\filldraw[black] (-1,1) circle (2pt) ;
\filldraw[black] (1,-1) circle (2pt) ;
\end{tikzpicture}
\caption{Diamond illustrating the nested structure of SL(2,$\mathbb{C}$) primary descendants that arises for primary operators with weights $(h,\bar h)=(\frac{1-k}{2},\frac{1-\bar k}{2})$ with $h,\bar k \in \mathbb{Z}_{>0}$.
}
\label{Nested_submodules}
\end{center}
\end{figure}
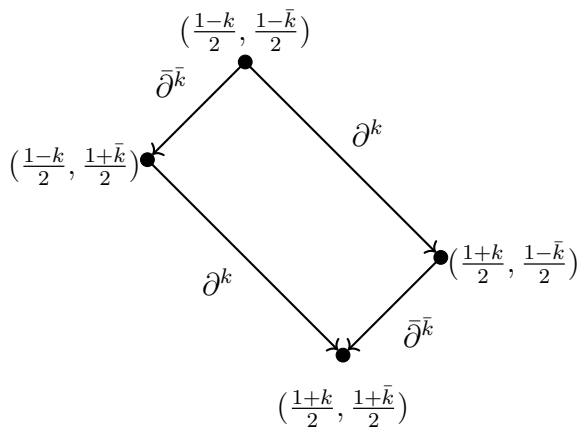
Let us start by considering the conditions for global primary descendants in generic two-dimensional CFTs.
An SL(2,$\mathbb{C}$) primary state $|h,\bar h \rangle$ 
is by definition annihilated by both $L_1$ and $\bar{L}_1$. We can construct its SL(2,$\mathbb{C}$) descendants by acting with $L_{-1}$ and $\bar{L}_{-1}$, or equivalently $\partial\equiv\partial_z$ and $\bar \partial \equiv \partial_{\bar z}$, an arbitrary number of times.
Focusing on the holomorphic algebra, a state of the form $(L_{-1})^k |h,\bar h \rangle $ is called a primary descendant if it is annihilated by $L_{1}$. This is satisfied for $h=\frac{1-k}{2}$ for $k \in \mathbb{Z}_{>0}$. The primary descendant has dimension $\frac{1+k}{2}$ corresponding to a reflection of the weight $h\to 1-h$. A similar reasoning can be repeated for the antiholomorphic algebra, yielding a primary descendant for ${\bar h}=\frac{1- \bar k}{2}$ for $\bar k \in \mathbb{Z}_{>0}$. When both conditions are satisfied, $(h,\bar{h})=(\frac{1-k}{2},\frac{1-\bar{k}}{2})$ the two submodules intersect at the position of an additional primary descendant forming a nested diamond structure.
This is illustrated in~Figure~\ref{Nested_submodules}. 

The conformally soft primaries in CCFT and their primary descendants are organized into {\it celestial diamonds} for which we now highlight some key aspects. The soft charge corresponding to a given asymptotic symmetry transformation can be expressed as~\cite{Banerjee:2018fgd,Banerjee:2019aoy,Banerjee:2019tam}
\begin{equation}\label{eq:QSdiamond}
 Q_S=\int d^2z \,\zeta(z,\bz)\cdot \calO_{soft}\,,
\end{equation}
where $\zeta$ is the symmetry transformation parameter and the~$\cdot$ takes care of tensor (or spinor) contractions. 
The operators $\calO_{soft}$ are primary descendants of conformally soft radiative fields ($J=\pm s$) and thus reside at the bottom corners of the celestial diamonds associated to $\zeta$. The two-dimensional Ward identities that are equal to four-dimensional soft theorems arise from the insertion of celestial currents $j,\bar j$ such that
\begin{equation}
    \bar \partial j=\calO_{soft}\,, \quad \partial \bar{j}=\calO_{soft}.
\end{equation}
This includes the U(1) current $J_z$ and the stress tensor $T_{zz}$ which are the conformally soft primary operators at the right corner of the diamond associated to, respectively, the leading photon and the sub-leading graviton soft theorem. The supertranslation current $P_z$ corresponds to the (non-primary) descendant of the primary operator at the right corner of the leading soft graviton diamond.  

A classification of all global conformal multiplets relevant in two-dimensional CCFT is given in~\cite{Pasterski:2021fjn}. We distinguish three types of primary descendants depending on whether the spin $J$ of the primary is bigger, equal or smaller in magnitude than that of the primary descendant. We can group them in the following three categories according to their ranges of $\Delta\in \mathbb{Z}$ focusing on integer spin.
\begin{itemize}
    \item[$\diamond$] {$\boldsymbol{1-|J|<\Delta<1+|J|}$}: The operator $\mathcal{O}_{\Delta,J}$ gives the soft charge~\eqref{eq:QSdiamond} for a corresponding asymptotic symmetry whenever the wavefunction $\Phi_{\Delta,J}$ is pure gauge as is the case for $1-|J| < \Delta \le 1 $. These conformally soft operators lie at the left or right corners of the celestial diamond and give rise to the leading (leading and sub-leading) soft theorem in gauge theory (gravity) and their associated memory effects. The remaining modes at $1 \le \Delta < 1+|J| $ in this range give rise to symplectically paired Goldstone operators which encode conformal Faddeev-Kulish dressings for celestial amplitudes.~\cite{Pasterski:2021dqe}
    \item[$\diamond$] $\boldsymbol{\Delta=1-|J|}$:  For the sub-leading (sub-sub-leading) soft theorem in gauge theory (gravity)~\cite{Low:1954kd,Cachazo:2014fwa} the diamonds degenerate to a line, and the primary operator with conformal dimension $\Delta=1-|J|$ descends to its own shadow with dimension $1+|J|$. Charges that give rise to an isomorphism with the soft theorems were discussed in~\cite{Donnay:2022sdg} while combinations of positive and negative spin modes make contact to the overleading large gauge transformations of~\cite{Campiglia:2016hvg,Campiglia:2016efb}.
    \item[$\diamond$] $\boldsymbol{\Delta<1-|J|}$: There are infinitely many primaries at $\Delta=1-|J|-n$ for $n>0$ whose descendant wavefunctions at level~$n$ naively vanish~\cite{Pasterski:2021fjn,Donnay:2022sdg}, while the algebra of primary operators constructed as in section~\ref{ssec:winf} is anything but trivial~\cite{Guevara:2021abz, Himwich:2021dau}. The associated charges generating the tower of soft theorems were identified in \cite{Freidel:2021ytz}. 
\end{itemize}
The above towers of conformally soft primary operators in gauge theory and gravity obey the infinite-dimensional holographic symmetry algebras discussed above.

\subsection{Conformal Dressings}
\label{subsec:dressing}

In QED and gravity infrared divergences exponentiate and set all matrix elements to zero. All-loop amplitudes admit a splitting into a soft and a hard component of the form \cite{Weinberg:1965nx}
\begin{equation} 
\label{eq:s-h-fact}
\mathcal{A} = e^{B}\mathcal{A}_{\rm bare},
\end{equation}
where in QED
\begin{equation} 
B = -\alpha \sum_{i < j} Q_i Q_j \ln \left|\frac{1}{2} p_i \cdot p_j\right|,
\end{equation}
whereas in gravity
\begin{equation}
B = -\gamma \sum_{i , j} (p_i \cdot p_j) \ln(p_i \cdot p_j) .
\end{equation}
Here $\alpha = \frac{e^2}{4\pi^2} \ln \Lambda_{IR}$ and $\gamma = \frac{G}{\pi}\ln \Lambda_{IR}$ with $\Lambda_{IR}$ an infrared cut-off, while $\mathcal{A}_{\rm bare}$ is the infrared-finite part of the amplitude which in our conventions is independent of both IR and UV cutoffs. 

In a conformal primary basis a similar decomposition continues to hold. This is not a priori obvious since both $B$ and $\mathcal{A}_{\rm bare}$ depend on the external energies and the Mellin integrals could in principle spoil factorization. Nevertheless, the celestial amplitudes derived from \eqref{eq:s-h-fact} take the form \cite{Arkani-Hamed:2020gyp}
\begin{equation}
\label{eq:conf-soft-fact}
    \widetilde{\mathcal{A}} = \widetilde{\mathcal{A}}_{\rm soft} \widetilde{\mathcal{A}}_{\rm hard}.
\end{equation}
In (massless) QED 
\begin{equation}
\begin{split}
    \widetilde{\mathcal{A}}_{\rm soft} \equiv \prod_{i < j} \left( z_{ij} \bar{z}_{ij} \right)^{-\alpha Q_i Q_j}, \qquad
    \widetilde{\mathcal{A}}_{\rm hard} \equiv \prod_i \int_0^{\infty} d\omega_i \omega_i^{\Delta_i' - 1}\mathcal{A}_{\rm bare}, 
    \end{split}
\end{equation}
with $ \Delta_i' = \Delta_i + \alpha Q_i^2$, while in gravity 
\begin{equation}
\begin{split}
    \widetilde{\mathcal{A}}_{\rm soft} \equiv \exp{\left[\sum_{i,j} 2\gamma P_i P_j |z_{ij}|^2 \log|z_{ij}|^2\right]}  ,\qquad
    \widetilde{\mathcal{A}}_{\rm hard}  \equiv \prod_i \int_0^{\infty} d\omega_i \omega_i^{\Delta_i - 1} \mathcal{A}_{\rm bare},
    \end{split}
\end{equation}
where $P_i$ should be understood as $P_{-\frac{1}{2}, -\frac{1}{2}}$ acting on the $i$-th leg according to \eqref{eq:ta}.

In both QED and gravity, the soft component of the celestial amplitude can be reproduced by vertex operators of Goldstone bosons for large gauge symmetry and supertranslations respectively \cite{Nande:2017dba, Himwich:2019qmj, Himwich:2020rro}. In QED, the soft component is manifestly conformally covariant and the shifts in the dimensions of the hard part ensure the net celestial amplitude transforms as a conformal correlator of dimension $\Delta_i$ operators as it should by construction.  In gravity, the charges $Q_k$ are replaced by energies which in a conformal primary basis results in an operator valued soft component in \eqref{eq:conf-soft-fact}. 

The conformally soft-hard factorization \eqref{eq:conf-soft-fact} leads to a natural prescription of defining infrared-finite celestial amplitudes. One can dress celestial amplitudes by (hermitian conjugates of) the vertex operators of the Goldstone bosons responsible for $\widetilde{\mathcal{A}}_{\rm soft}$. In QED and gravity, the Goldstone bosons $ \mathcal S$ and $ \mathcal C$ are defined by 
\begin{equation}\label{eq:SC} 
 \mathcal S_{z} = i \partial_{z}  \mathcal S \,, \quad  \mathcal C_{zz} =i{\textstyle \frac{1}{2!}} \partial_{z}^2  \mathcal C
\end{equation}
where $\mathcal S_{z}$ and $ \mathcal C_{zz}$ are $\Delta=1$ operators defined by inner product~\eqref{eq:2Dop} of bulk operators of respectively spin $s=1$ and $s=2$ with conformal primary wavefunctions $A^{\rm CS}_{1,-1}$ and $h^{\rm CS}_{1,-2}$ that are canonically conjugate to the $\Delta=1$ Goldstone wavefunctions~\eqref{eq:A1} and~\eqref{eq:h1}, i.e. $i(A^{\rm CS},A^{\rm G})=i(h^{\rm CS},h^{\rm G})=(2\pi)^2\delta^{(2)}(z-z')$~\cite{Donnay:2018neh,Arkani-Hamed:2020gyp}. 
In QED, the vertex operators $e^{-iQ_k \mathcal S(z_k, \bar{z}_k)}$ can be shown to precisely coincide with the Lorentz-invariant Faddeev-Kulish dressings~\cite{Chung:1965zza, Kibble:1968lka, Kibble:1968npb, Kibble:1968oug, Kulish:1970ut} allowing for coherent photons of all energies. In a momentum space basis, momentum conservation induces an upper bound on the allowed energies of photons in the cloud. In a conformal primary basis such a constraint is irrelevant and a natural choice of the dressing is simply picked out by conformal symmetry.
Analogous statements hold for gravity and are worked out in~\cite{Arkani-Hamed:2020gyp}.

\section{Outlook}
\label{sec:open}
The importance of soft factorisation for the cancellation of IR divergences in QCD cross-sections ensures the continuing relevance of developing more efficient methods for perturbative computations. Additionally, the recent developments in applying amplitude methods to gravitational wave physics provides motivation for extending our current understanding of soft limits to new regimes such as  states with classical spin. Even at a more formal level there are open questions regarding the connection between soft theorems and known asymptotic symmetries, for example it would be useful to have a complete one-loop matching of the sub-leading soft-factor in gravity and the two-dimensional stress tensor Ward identity. 

The reformulation of the scattering problem in a conformal primary basis has so far proven particularly useful in identifying new symmetries of Nature. Indeed, the (semi-)infinite tower of tree-level symmetry currents discussed in the previous section is perhaps surprising from a momentum space point of view: we are only starting to uncover the interpretation and implications of these symmetries beyond sub-leading order! It seems likely they, as well as their extension beyond tree level, will imply powerful constraints on consistent low energy effective field theories.

Massless low-point celestial amplitudes, while conformally covariant, are plagued with singularities inherited from bulk momentum conservation. This fundamental difference to standard CFTs where such singularities are typically absent has led to challenges in applying CFT methods in the study of CCFTs. One promising way around this is the proposal that the conformal primary solutions should be traded for light- or shadow-transforms thereof~\cite{Pasterski:2017kqt,Crawley:2021ivb,Atanasov:2021cje,Sharma:2021gcz,Hu:2022syq} in which case contact terms may disappear. Nevertheless, whether there exists a basis in which all singular conformal structures become regular remains an important open question. 

Massive celestial amplitudes are currently poorly understood and deserve further study. While some of their symmetries were discussed in~\cite{Law:2019glh,Law:2020tsg}, there is evidence that massive operators are non-local on the celestial sphere~\cite{Kapec:2021eug} which obscures the power of celestial symmetries and complicates the application of the methods discussed in section~\ref{sec:Celestial-symmetries} to this case. A better understanding of massive celestial amplitudes would be especially interesting as it could provide insights into the holographic description of non-perturbative asymptotically flat backgrounds, including black holes beyond their ultraboosted limits~\cite{Pasterski:2020pdk}. 

Although recent progress was made in~\cite{Chang:2021wvv}, we are currently lacking an understanding of basic aspects of the holographic correspondence, such as the precise ways in which bulk locality, unitarity and causality are encoded in CCFTs. Moreover, there are subtleties involving OPE associativity in mixed helicity sectors. A resolution of these issues would pave the way towards setting up a bootstrap program in this context. Preliminary steps in this direction were taken in~\cite{Nandan:2019jas,Fan:2021isc,Atanasov:2021cje,Fan:2021pbp,Fan:2022vbz} where it was shown in a range of examples that celestial four-point amplitudes encode information about the spectrum and OPE coefficients of CCFTs not too unlike conventional CFTs. Identifying the central charge of CCFT would also be essential for the holographic dictionary.

An important milestone in the celestial holography programme is to identify an intrinsic CFT construction of a bulk theory. In the conformally soft sector, progress has been made in a series of recent works~\cite{Nande:2017dba,Magnea:2021fvy,Gonzalez:2021dxw,Cheung:2016iub,Nguyen:2020waf,Nguyen:2020hot,Kalyanapuram:2020epb,Kalyanapuram:2021bvf,Pasterski:2021dqe}  where effective actions have been proposed that capture infrared aspects of gauge theory and gravity both in four and more spacetime dimensions.

Since celestial amplitudes are expressed in boost eigenstates which superpose all energies, the usual Wilsonian decoupling of UV/IR physics no longer applies. Being well-defined only for theories that are equipped with a UV completion, celestial amplitudes offer an arena to study general properties of consistent quantum gravity theories. Interestingly, the study of celestial superstring amplitudes~\cite{Stieberger:2018edy} suggests that the string world sheet becomes celestial in a certain limit. The embedding of celestial holography in string theory is an exciting open direction.

\section*{Acknowledgments}
We are grateful to our collaborators and colleagues with whom we enjoyed many discussions and projects on aspects of the topic of this review.
This work was supported by the European Union's Horizon 2020 research
and innovation programme under the Marie Sk\l{}odowska-Curie grant
agreement No.~764850 {\it ``\href{https://sagex.org}{SAGEX}''}. TMcL is supported by Science Foundation Ireland through grant grant 15/CDA/3472. AP is supported by the European Research Council (ERC) under the European Union’s Horizon 2020 research and innovation programme (grant agreement No 852386). A.R. is supported by the Stephen Hawking fellowship. Research at
Perimeter Institute is supported in part by the Government of Canada through the Department of Innovation, Science and Economic Development Canada and by the Province of Ontario through the Ministry of Colleges and Universities. 

\section*{Bibliography}
\bibliographystyle{iopart-num}
\newcommand{\eprint}[2][]{\href{https://arxiv.org/abs/#2}{\tt{#2}}}
\bibliography{references.bib}

\end{document}